\documentclass[journal]{vgtc}                     % final (journal style)
\onlineid{1351}

\vgtccategory{Research}

\vgtcpapertype{application/design study}

\title{Sportify: Question Answering with Embedded Visualizations and Personified Narratives for Sports Video}

\author{%
  Chunggi Lee,
  Tica Lin,
  Hanspeter Pfister,
  and Chen Zhu-Tian
}

\authorfooter{
  \item Chunggi Lee, Tica Lin, and Hanspeter Pfister are with John
A. Paulson School of Engineering and Applied Sciences, Harvard University,
Cambridge, MA. E-mail: {chunggi\_lee, mlin, pfister}@g.harvard.edu

  \item Chen Zhu-Tian is with CSE department, University of Minnesota, Minneapolis, MN. E-mail: ztchen@umn.edu. The work was partially done when he was with Harvard.
}

\abstract{As basketball's popularity surges, fans often find themselves confused and overwhelmed by the rapid game pace and complexity. Basketball tactics, involving a complex series of actions, require substantial knowledge to be fully understood. This complexity leads to a need for additional information and explanation, which can distract fans from the game. 
To tackle these challenges, we present \ourtool{}, a Visual Question Answering system that integrates narratives and embedded visualization for demystifying basketball tactical questions, aiding fans in understanding various game aspects. 
We propose three novel action visualizations (i.e., Pass, Cut, and Screen) to demonstrate critical action sequences. To explain the reasoning and logic behind players' actions, we leverage a large-language model (LLM) to generate narratives. We adopt a storytelling approach for complex scenarios from both first and third-person perspectives, integrating action visualizations. We evaluated \ourtool{}  with basketball fans to investigate its impact on understanding of tactics, and how different personal perspectives of narratives impact the understanding of complex tactic with action visualizations. 
Our evaluation with basketball fans demonstrates \ourtool{}'s capability to deepen tactical insights and amplify the viewing experience. 
Furthermore, third-person narration assists people in getting in-depth game explanations while first-person narration enhances fans’ game engagement.
}


\teaser{
  \centering
  \includegraphics[width=\linewidth, alt={A view of a city with buildings peeking out of the clouds.}]{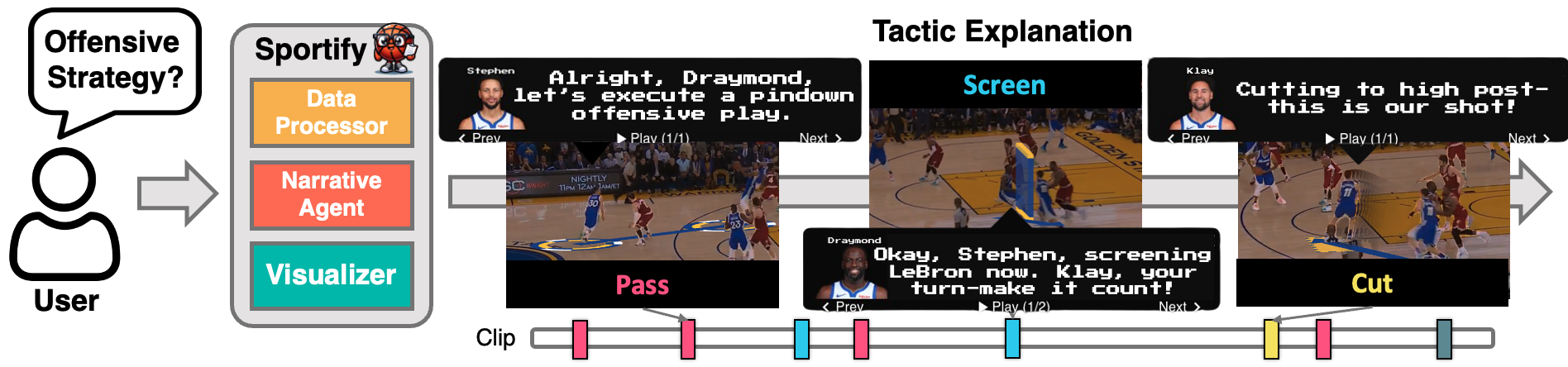}
  \caption{%
  	\ourtool{}  explains tactic questions in each clip for everyone, aiming to engage users and foster a love for sports. We integrate embedded visualization and personified narratives generated by large language model (LLM) to elucidate a complex series of actions through action detection, tactic classifier, and LLM pipelines.%
  }
  \label{fig:teaser}
}

\graphicspath{{figs/}{figures/}{pictures/}{images/}{./}} % where to search for the images

\usepackage{tabu}                      % only used for the table example
\usepackage{booktabs}                  % only used for the table example
\usepackage{lipsum}                    % used to generate placeholder text
\usepackage{mwe}                       % used to generate placeholder figures

\usepackage{mathptmx}                  % use matching math font
\usepackage{enumitem}

\usepackage[scaled]{beramono}
\usepackage[T1]{fontenc}
\usepackage{stfloats}
\usepackage{array}
\usepackage{adjustbox}
\usepackage{listings}

\begin{document}

%%%%%%%%%%%%%%%%%%%%%%%%%%%%%%%%%%%%%%%%%%%%%%%%%%%%%%%%%%%%%%%%
%%%%%%%%%%%%%%%%%%%%%% START OF THE PAPER %%%%%%%%%%%%%%%%%%%%%%
%%%%%%%%%%%%%%%%%%%%%%%%%%%%%%%%%%%%%%%%%%%%%%%%%%%%%%%%%%%%%%%%

\newcommand{\revision}[1]{\textcolor{black}{#1}}
\newcommand{\ourtool}{Sportify}
\newcommand{\para}[1]{\vspace{1mm}\noindent\textbf{#1}}
\newcommand{\zt}[1]{{\color{blue}{[ZT: #1]}}}
\newcommand{\Cut}{\textsc{Cut}}
\newcommand{\Pass}{\textsc{Pass}}
\newcommand{\Shoot}{\textsc{Shoot}}
\newcommand{\Screen}{\textsc{Screen}}
\newsavebox{\coloredquotationbox}
\newenvironment{coloredquotation}
 {%
  \begin{trivlist}
  \item\relax
  \begin{lrbox}{\coloredquotationbox}
  \setlength{\fboxsep}{10pt} 
  \begin{minipage}{\dimexpr\linewidth-2\fboxsep}
 }
 {%
  \end{minipage}
  \end{lrbox}
  \item\relax
  \parbox{\linewidth}{
    \begingroup
    \color[RGB]{224,215,188}%
    \hrule
    \color[RGB]{249,245,233}%
    \hrule
    \color[RGB]{224,215,188}%
    \hrule
    \endgroup
    
     % \par\noindent % Ensure no indentation at the beginning of the box
    \setlength{\fboxsep}{10pt} % Adjust padding around the content inside the box
    \colorbox[RGB]{249,245,233}{\usebox{\coloredquotationbox}}\par\nointerlineskip
    
    \begingroup
    \color[RGB]{224,215,188}%
    \hrule
    \color[RGB]{249,245,233}%
    \hrule
    \color[RGB]{224,215,188}%
    \hrule
    \endgroup
  }
  \end{trivlist}
 }
\newsavebox{\coloredquotationbluebox}
\newenvironment{coloredquotationblue}
{%
\begin{trivlist}
\item\relax
\begin{lrbox}{\coloredquotationbluebox}
\setlength{\fboxsep}{10pt} 
\begin{minipage}{\dimexpr\linewidth-2\fboxsep}
}
{%
\end{minipage}
\end{lrbox}
\item\relax
\parbox{\linewidth}{
\begingroup
\color[RGB]{136, 157, 181}%
\hrule
\color[RGB]{212,224,238}%
\hrule
\color[RGB]{136, 157, 181}%
\hrule
\endgroup

 % \par\noindent % Ensure no indentation at the beginning of the box
\setlength{\fboxsep}{10pt} % Adjust padding around the content inside the box
\colorbox[RGB]{212,224,238}{\usebox{\coloredquotationbluebox}}\par\nointerlineskip

\begingroup
\color[RGB]{136, 157, 181}%
\hrule
\color[RGB]{212,224,238}%
\hrule
\color[RGB]{136, 157, 181}%
\hrule
\endgroup
}
\end{trivlist}
}

%% The ``\maketitle'' command must be the first command after the
%% ``\begin{document}'' command. It prepares and prints the title block.
%% the only exception to this rule is the \firstsection command
\firstsection{Introduction}
\maketitle
Basketball attracts 400 millions fans worldwide~\cite{popularSports}, with the NBA Finals alone drawing a peak audience of 17 million~\cite{nbaviewership}. 
Despite widespread interest in basketball, the rapid pace and intricate dynamics of the basketball plays often leave fans confused and eager for a deeper understanding of the game~\cite{chen2023iball, lin2022quest}.
Commentary, while helpful, often lacks in providing the abundance of information fans desire, from player performance metrics to complex tactical decisions. 
Particularly, understanding the \emph{tactics} represents a significant but challenging task. 
It involves \emph{a series of actions}—screening, passing, cutting, and shooting—that requires significant knowledge to be fully understood~\cite{sicilia2019deephoops}.
These tactics are crucial in maximizing scoring opportunities, from single plays to team-wide tactics~\cite{kohli2015optimal, singh2023optimizing, mcintyre2016recognizing}.
Players constantly make split-second decisions to either take the shot or distribute the ball, optimizing their team's offensive tactics~\cite{skinner2017optimal}. 
A deeper comprehension of these tactics not only enhances the watching experience but also deepens fans' engagement of the game.

Recent efforts in bridging the knowledge gap for fans have embraced the concept of \emph{embedded visualizations}~\cite{willett2016embedded}. 
These visualizations 
enrich the watching experience by seamlessly integrating 
data insights within the physical context of the game.
Therefore, 
embedded visualizations have been widely used by both commercial ~\cite{secondspectrum, courtvision, vizlibero} and research systems~\cite{Chen2021Augmenting, zhu2022sporthesia, chen2023iball, lin2022quest} to create \emph{augmented sports videos}.
Current augmented sports video products and systems are innovative but limited to deploying predefined visualizations, 
offering a fixed set of insights for the game's inherently dynamic character~\cite{chen2023iball, lin2022quest}. 
Notably, these tools lack interactive features that would allow fans to investigate the game's tactical dimensions, a limitation that restricts personalized engagement and understanding of complex tactics~\cite{lin2022quest}.
This reveals the requirement for advanced game watching systems that enable fans to dynamically explore and query game tactics.

In this paper, we develop a novel Visual Question Answering (VQA) system, \ourtool{} (\autoref{fig:teaser}), 
which integrates embedded visualization and personified narratives to explain basketball tactics.
\revision{As a VQA system, \ourtool{} focuses on understanding and answering questions about input images, rather than generating visual answers~\cite{de2023visual}.}
% In this paper, we present \ourtool{} (\autoref{fig
% }), a Visual Question Answering (VQA) system that uses embedded visualization and personified narratives to explain basketball tactics. Our VQA system focuses on understanding input images rather than generating visual answers \cite{de2023visual}.
The design of \ourtool{} is guided by three design considerations 
thereby enhancing the explanations it provides:
1) video-based tactic explanations,
2) narrative tactic explanations, 
and 3) embedded visual tactic explanations.
Together, these design considerations are crafted to ensure that
the generated tactic explanations are reliable, understandable, and engaging for the users.

% \textbf{in the form of augmented sports videos}. 
To answer questions about tactics, % and in-game decisions, 
\ourtool{} leverages a machine learning pipeline to identify four actions (i.e., \Pass{}, \Cut{}, \Screen{}, and \Shoot{}),
which are critical to understanding basketball tactics,
by leveraging the on-court positions of the players.
% by utilizing player and ball coordinates and bounding boxes. 
A K-Nearest Neighbors (KNN) algorithm classifies the tactics based on the players' coordinates. 
The results are subsequently used to generate answers 
about the logic behind players' actions 
by using a large language model (LLM).
To present the answers, 
we propose three new action visualizations for three key actions 
(i.e., \Pass{}, \Cut{}, and \Screen{}) 
and two personified narratives (i.e., \textbf{first-person} and \textbf{third-person} perspectives), 
explaining the tactics in a storytelling format.
% Through the integration of embedded visualization and narrative, 
By this, \ourtool{} transforms the passive viewing into an active exploration experiences, 
offering a deeper understanding and engagement with the game.

We conducted two user studies (i.e., a comparative study and a exploratory study) with basketball fans to investigate the impact on understanding of tactics and in-game decisions.
We compare three different conditions and two narrative perspective to explore their impacts on the comprehension of basketball tactics.
% with action visualizations. 
The results demonstrate that \ourtool{} helps users understand tactics and in-game decisions better, 
and enhances user engagement and experience compared to existing tactic explanation videos. 
Furthermore, third-person perspective narration assists people in getting in-depth game explanations while first-person narration enhances fans' game fan and engagement.

In summary, our contributions are as follow: 1) the implementation and design of a novel VQA system for answering tactics and in-game decisions, 2) three novel visualizations such as Cut, Screen, and Pass, integrating the personified narratives (i.e., first or third-person perspective) to provide a storytelling experience that enhances the engagement for fans, and 3) two user studies that evaluate three different conditions with the two personified perspectives texts, and the usability of \ourtool{} comparing to existing tactic explanation videos.

\section{Related Work}
\subsection{Visual Analytics in Sports}
Sports data are intrinsically spatial and dynamic. To support analyzing complex sports data for enhancing game understanding and performance evaluation, visualization researchers have developed novel visualization techniques for different sports data and target users. 

In particular, basketball has inspired a plethora of novel visualization designs with its intricate interplay among team members and the detailed, organized spatiotemporal data. 
Targeting sports analysts, BKViz~\cite{Losada2016BKVizAB} designed an interactive visual analytic system for analyzing individual player performance and team dynamics in a basketball game. Users can analyze heterogeneous data with novel visualizations to support finding patterns and correlations between player actions and performance, such as play-by-play data on a court diagram and player interaction in an arc diagram.
OBTracker~\cite{Wu2022OBTrackerVA} focused on evaluating the contribution of the player's off-ball movement and presented the player action type and performance in a novel glyph and Voronoi diagram design.
HoopInSight~\cite{Fu2023HoopInSightAA} compared players' shooting performances using side-by-side shot heat maps and aggregated spatial data presented as location-based glyphs. 
Other popular sports also attract much attention in the visualization community, including soccer~\cite{Perin2013SoccerStoriesAK,Stein2018BringIT}, baseball~\cite{Dietrich2014Baseball4DAT}, table tennis~\cite{Wang2021TacMinerVT,Wang2020TacSimurTS}, and badminton~\cite{Chu2021TIVEEVE, Lin2023VIRDIM}. These studies contribute novel visualization approaches to enhance spatiotemporal data analysis and communications in their respective sports.
 
Targeting non-data experts, some work focused on novel interaction and visualization techniques to support seamless analytic workflow. To support analyzing data during live game viewing,
GameViews~\cite{Zhi2019GameViewsUA} presents box score views and a game flow chart with key events, along with a chat feature for basketball fans to analyze and discuss game insights live.
GameBot~\cite{Zhi2020GameBotAV} proposed using the conversational interface for fans to retrieve game-related data during live basketball games instantly and designed mobile visualizations to enhance data understanding.
Lin et al.~\cite{lin2022quest} proposed an embedded visualization framework for analyzing game data within the game context in the basketball game view without the need to switch contexts. 
More recently, immersive technologies were used to enhance interactive data analysis for coaches and players in racket sports. 
TIVEE~\cite{Chu2021TIVEEVE} designed an interactive VR interface with embodied interaction to allow analyzing badminton trajectory data in an overview of small multiples and a live-sized badminton court view. VIRD~\cite{Lin2023VIRDIM} further developed a 3D reconstructed game view based on 2D badminton game videos to support deeper insight analysis for high-performance coaching.
Our study builds upon the rich work in sports visualizations and develops novel action-based visualizations for game tactics and in-game decision-making in sports videos.

\subsection{Embedded Visualization in Sports Videos}
With advanced computer vision techniques, recent research focused on designing visualizations that are directly embedded into sports videos to enhance game analysis of dynamic sports movement.

Stein et al.~\cite{Stein2018BringIT} developed a visual analytic system that combines soccer game videos with trajectory visualizations, applying computer vision methods to derive trajectory measures from the video inputs. Their results show that this embedded visualization method enables expert analysts to perform effective contextualized analysis on team performance. 
Zhu-Tian et al.~\cite{Chen2021Augmenting} proposed a direct manipulation user interface to allow a direct link of the game data to a selected player and contributed a design framework for embedding visual elements with video effects into sports videos, which supports presenting data insights in the sports videos effectively. 
Lin et al.~\cite{lin2022quest} further tackled the problem of dynamic data requirements throughout sports games by proposing a context-driven embedded visualization framework for live game analysis of sports fans.
Yao et al.~\cite{Yao2023DesigningFV} examined the challenges and design considerations for embedded visualizations in swimming videos from the designer's perspective and found motion context has an impact on the visualization design choices.
Zhu-Tian et al.~\cite{chen2023iball} designed gaze interaction to moderate the visualizations shown in basketball videos to avoid visual clutter and enhance fans' game understanding with adaptive visualizations.

% With increasing interest in embedding data through visualizations in sports videos, recent work investigates interactions that support users in consuming and analyzing these embedded visualizations in sports videos. 
% Lin et al.~\cite{lin2022quest} explored using voice commands for activating different embedded visualizations during live game viewing. They found fans preferred different interaction strategies based on individual preferences and knowledge, where some preferred predefined sets of visualizations and some preferred active interaction during the game.
% Chen et al.~\cite{chen2023iball} designed gaze interaction to moderate the visualizations shown in basketball videos to avoid visual clutter and enhance fans' game understanding with adaptive visualizations. Their user study has found that gaze interactions encouraged data exploration during game viewing and that some fans requested more customization to retrieve data they are interested in.

% 
Based on the prior research, we identified two gaps in utilizing embedded visualizations within sports videos to enhance fans' game comprehension. First, existing studies predominantly focus on presenting metadata and game statistics (such as athlete names and speeds in~\cite{Yao2023DesigningFV}) and spatial information (like trajectories and zones in~\cite{Stein2018BringIT}). Aspects that involve more complex data, like game tactics and in-game decision-making, were only preliminary explored by Zhu-Tian et al.~\cite{Chen2021Augmenting}, which did not target sports fans. 
Second, there is a lack of customized interactions to support users in retrieving data and analyzing game context during the game. 
Present approaches tend to offer limited engagement options, such as passive viewing (e.g., gaze in~\cite{chen2023iball}) or basic voice commands~\cite{lin2022quest}, which fall short of meeting fans' expectations and allowing more complex data to be explored.
Our study addresses this gap by proposing a visual question-answering system that allows fans to explore complex game tactics within sports videos through active conversation and embedded visualizations tailored for action-rich game contexts.

\begin{figure*}[t!]
\centering
\includegraphics[width=1.0\textwidth]{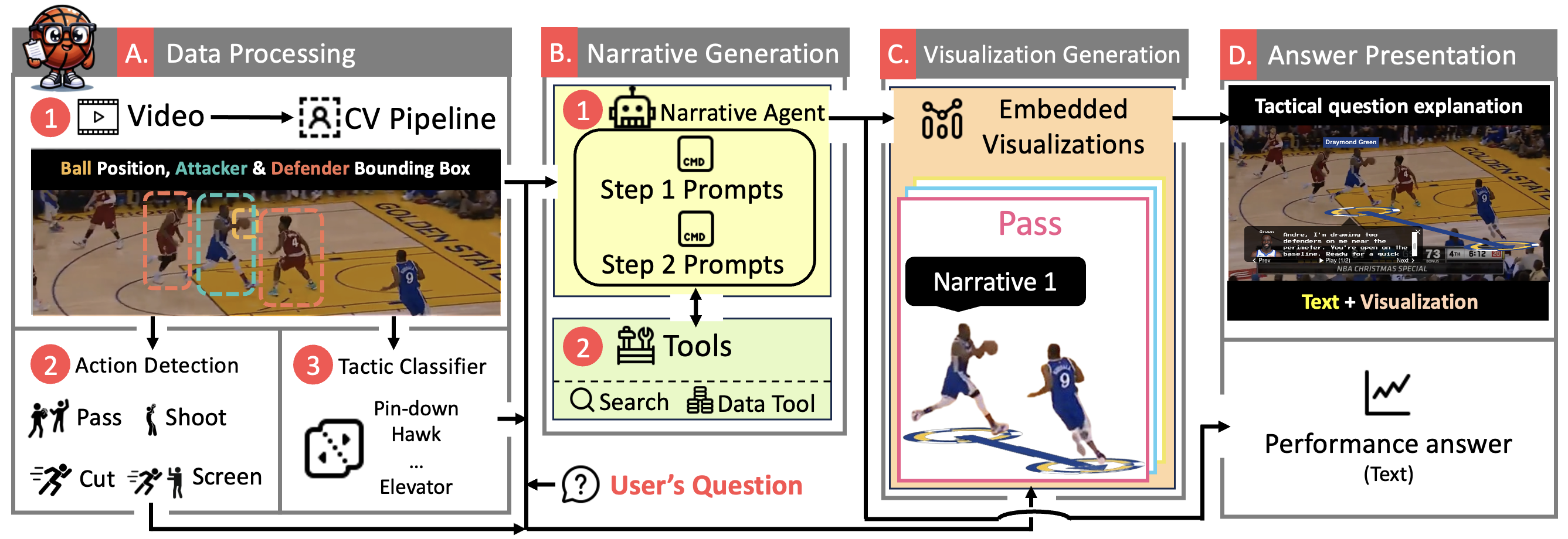}
\caption{The pipeline efficiently addresses both tactical and performance-based questions. It begins with data processing (A-1), where videos undergo a computer vision (CV) pipeline to identify players' coordinates, bounding boxes, and the ball's location. This information feeds into action detection (A-2) and tactic classification (A-3), generating tactical textual information for the narrative agent (B-1). Player coordinates and LLM responses are visually embedded (C) and displayed in the video (D). Performance-related queries are handled by the LLM, which retrieves data to provide text-based answers (D).}
\label{fig:system}
\vspace{-0.5cm}
\end{figure*}

\subsection{Visual Storytelling for Spatiotemporal Data}

Visual storytelling, especially in the context of spatiotemporal data, has proven to be an effective tactic for communicating complex data~\cite{segel2010narrative}.
Research grounded in cognitive science has demonstrated that integrating visual and verbal elements enables the construction of a cohesive mental model of a narrative, thereby enhancing the comprehension of complex data~\cite{schnotz2005integrated}.
Mayr et al.~\cite{mayr2018once} investigated the organization of temporal and spatial information in supporting narrative comprehension and identified five hybrid visualization techniques, including multiple coordinated views~\cite{roberts2005exploratory}, animations~\cite{kriglstein2014pep}, layer superimposition~\cite{javed2012exploring}, layer juxtaposition (or data comics~\cite{zhao2015data}), and space-time cube~\cite{kraak2003space}.
% 
% Mayr et al.~\cite{mayr2018once} investigated the human internal representation of visual storytelling of spatiotemporal data from the cognitive science perspective. They found that visual and verbal information are processed in parallel and integrated to construct a coherent mental model of narrative, grounded in Schnotz's integrated model of human comprehension~\cite{schnotz2005integrated}. Furthermore,
% temporal and spatial information as two key dimensions in a story  
% require distinct organization of visual space (e.g., timeline or map), which demands hybrid visualization techniques to support comprehension, including multiple coordinated views~\cite{roberts2005exploratory}, animations~\cite{kriglstein2014pep}, layer superimposition~\cite{javed2012exploring}, layer juxtaposition (or data comics~\cite{zhao2015data}), and space-time cube~\cite{kraak2003space}.
As each technique has strengths and limitations in conveying narrative data, it is important to make careful design choices to assist users' internal representation in linking this multimodal information. 
% Mayr et al.~\cite{} proposed a layout transition technique to connect multiple views seamlessly.

Drawing upon this visual storytelling framework to convey complex tactics and decisions in sports, two areas of research are of interest: visual representation of spatiotemporal data and verbal narrative techniques.  
Several hybrid visualization techniques were adopted for dynamic sports data,
including multiple coordinated views~\cite{Zhi2019GameViewsUA}, layer superimposition~\cite{Perin2013SoccerStoriesAK}, and animations~\cite{Chen2021Augmenting}. 
SoccerStories~\cite{Perin2013SoccerStoriesAK} visualizes the sequence of actions in a soccer play with trajectories and linked faceted views on a court diagram. They further proposed a small multiple technique that embeds these diagrams in sports articles to support journalists in communicating data insights in stories.
% GameViews~\cite{Zhi2019GameViewsUA} carefully designed multiple views of game data visualizations based on sportswriters' workflow, including linked charts to represent game statistics related to temporal and spatial information, allowing sportswriters to comprehend multimodal data and generate multimedia data stories effectively through interactions.
VisCommentator~\cite{Chen2021Augmenting} proposed a framework for embedding animated visualizations in sports videos, supporting the creation of data videos with narrative structures like linear and flashback. 
% Their system supported domain experts in creating animated data videos to explain spatiotemporal data insights.

In addition to organizing spatial and temporal data views, it is important to explain causality in the sequence of actions for sports game tactics. As shown by Choudhry et al.~\cite{choudhry2020once}, natural language narratives can complement visualizations when explaining complex causality in network data.  
Despite different data types, the causal sequences in sports actions share similarities. 
Zhu-Tian et al.~\cite{zhu2022sporthesia} integrated natural language commentary with visual animation in racket sports videos, allowing coupling narratives with animated embedded visualizations to explain game actions in more detail.
Building upon the existing work, our work explores using a natural language approach to construct data storytelling for complex sports tactics in basketball. Our novelty lies in adopting LLM-based question-answering techniques to create textual narratives, coupled with animated embedded visualizations in sports videos to create personalized visual storytelling. We also explore how different narrative perspectives, including first- and third-person, impact data comprehension and engagement. To the best of our knowledge, this is the first initiative to blend LLM-generated text narratives with question-answering for sports visual storytelling.

\section{Designing \ourtool{}}

This section first describes the design considerations behind \ourtool{}
and then overviews its three major components.

\subsection{Design Considerations}
Previous works~\cite{chen2023iball, lin2022quest} identified fans are curious and eager to understand the tactics and in-game decisions (e.g., \textit{``the usage of this particular play''} and \textit{``understand the offensive and defensive strategies''}).
Thus, our QA system specifically focuses on answering questions related to basketball tactics.
% However, the complexity of basketball tactics 
Yet, this presents significant challenges,
as the answers should \emph{explain a series of complex actions} 
and presenting the logic behind player movements in an \emph{reliable}, \emph{understandable}, and \emph{engaging} manner.
% Informed by previous research~\cite{},
We conceived the design considerations as follow:

\para{R1. Reliable -- Explaining Tactics with Grounded Video Data.}
Ensuring accuracy and reliability in explanations is a critical requirement for QA systems, particularly those analyzing video content. 
The alignment between the video content and the provided explanations is essential, as the mismatches between the video content and the provided explanations can lead to significant user confusion. 
This necessitates a mechanism to understand the video and extract data from the video, such as tactic types and the actions involved.
This data then serves as the foundation for the QA system to generate explanations.
By ensuring that our explanations are directly tied to the observable tactics and actions in the video, we provide users with insights that are not only precise but also verifiable, enhancing the reliability of the information presented by \ourtool{}.

% A notable challenge in utilizing LLMs for explanation generation is their their tendency for 
% ``hallucination''~\cite{} – the creation of information that is not factual.
% This tendency is especially problematic in video-based question answering systems, 
% where mismatch between the video content and the provided explanations can lead to user confusion. 
% To address this issue, our strategy prioritizes the accurate grounding of explanations within the actual video content. 
% We advocate for the use of external methods to extract data from the video, such as tactic types and the actions involved.
% This data then serves as the foundation for the LLM-generated explanations.
% By ensuring that our explanations are directly tied to the observable tactics and actions in the video, we provide users with insights that are not only precise but also verifiable, enhancing the reliability of the information presented by 
% \ourtool{}.

\para{R2. Understandable - Explaining Tactics with Narratives.}
Storytelling is fundamental in organizing and conveying human experiences, 
playing a crucial role in how we understand and interpret events~\cite{mayr2018once, schroder2023telling}.
To help users easily understand the tactics employed by teams,
we propose the use of well-structured \emph{narratives} that adhere to a logical sequence. 
This approach not only clarifies the sequence of actions 
but also reveals the underlying reasons and objectives guiding the players' movements~\cite{choudhry2020once}. 
Moreover, the choice of narrative perspective (i.e., first, second, or third person) also demands careful consideration, as it significantly impacts a viewer's engagement and immersion~\cite{chen2021changing}. 
In basketball, the third-person perspective aligns with a commentator's view, offering a broad overview of the game, while the first-person perspective resonates with the individual player's decision-making process. 
Each perspective offers potential benefits in game understanding.
Finally, it is essential that these narratives are not only presented as plain text 
but organized in a structural format that can be effectively mapped or linked with visualizations embedded within the video.

\para{R3. Engaging -- Explaining Tactics using Embedded Visualizations.} %% 3 Action Vis + dialouge bubble
To enrich the explanation of tactics, it is essential to complement the narratives with visual representations, creating an experience akin to watching a film~\cite{gershon2001storytelling}. 
This requires the careful design of embedded visualizations for the key actions within the narratives. 
Each action demands specific animated embedded visualizations
to capture its unique objectives and to dynamic nature.
Furthermore, the visual explanation must also reflect the chosen narrative perspective.
For instance, in a first-person narrative, 
the narration should directly connect to specific subjects within the visualization. 
In contrast, a third-person narrative allows for a more generalized correlation between the visualizations and the narration.
The ultimate challenge for \ourtool{} is to seamlessly integrate these visual explanations within the video content, ensuring a cohesive and engaging presentation of tactical explanation. 

\subsection{System Overview}

\noindent
Based on the considerations, we have developed \ourtool{}, 
a visual QA system \revision{answering questions about videos~\cite{de2023visual}  and} comprising three major components: 
a Data Processor (\autoref{fig:system} a), 
a Narrative Agent (\autoref{fig:system} b),
and a Visualizer (\autoref{fig:system} c). 

At the heart of \ourtool{} lies the Narrative Agent, which leverages a LLM to interpret the user's question and generate explanations in response.
For a system designed for basketball videos, the capability 
to understand video content is indispensable.
Although multi-modal LLMs are capable of processing image data, they often underperform in domain-specific tasks  and require a tremendous computation costs, such as detecting actions or tactics from a sports video. 
To overcome this challenge, our methodology employs a text-only LLM, enriched through the integration of a Retrieval-Augmented Generation (RAG) framework~\cite{lewis2020retrieval} 
and a Reasoning-and-Actioning (ReAct) prompting strategy~\cite{yao2022react}
for different types of questions.
Importantly, \ourtool{} leverages the data extracted from the video as the context (R1) 
to generate the explanation in a narrative format (R2).
These extracted data and explanation are then presented as visualizations embedded in the video (R3). 
In the subsequent sections, we delve into the specific design and implementation of each component.

% The forthcoming sections will provide a detailed exploration of the design and implementation of each component, laying the groundwork for understanding how \ourtool{} enhances the experience of interacting with basketball video content.

% The core of \ourtool{} is a LLM-based agent, so called Narrative Agent.
% Leveraging LLM provides a promising solution to accomplish these objectives.
% To fulfill R1, our system should be able to 1) accurately interpret users' natural language queries, and 2) generate insightful explanations in response.
% However, generating meaningful explanations requires the LLM to 

% and contextual data  to craft a comprehensive explanation. 

% Based on the design requirements analyzed in Section 3, we introduce \ourtool{}, a visual question-answering (VQA) system designed to elucidate complex questions (R1), such as those related to strategy and in-game decisions (R2). This tool combines narrative and visualization techniques for effective storytelling (R3). \ourtool{} consists of two pipeline parts that allow users to ask not only strategic and in-game decision questions but also performance- and stats-based questions.

\section{Data Processor}
\label{sec:data_processor}
To create reliable explanations (R1),  extracting data from video clips as context for the LLM is crucial.
We detail our data extraction method, focusing on identifying tactics and actions, by leveraging the 3D coordinates from the publicly available SportsVU dataset~\cite{SportVU24} and applying machine learning to determine players and ball positions, following prior work~\cite{chen2023iball}.
% This section outlines our approach to extract the data:
% identifying the tactics and actions involved. 
% We utilize 3D coordinates from  publicly available SportsVU dataset~\cite{SportVU24} and employ machine learning techniques to gain the players and ball bounding box in video like previous work~\cite{chen2023iball}.
% Following previous work~\cite{},
% we employ a suite of machine learning techniques to analyze the players' on-court coordinate data,
% which provides a precise player positions.
The 3D coordinates is tracked through multi-camera tracking systems \cite{sportvucamera}, a technique widely adopted in professional basketball leagues, including the NBA~\cite{nbawebsite}. 
In this study, we utilize the 3D coordinates and extracted data from videos to identify tactics and actions for effectiveness and applicability.

% from SportVU \cite{SportVU24} and predict the label of tactics, as shown in \autoref{fig:system} (A-3).
% Our solution is to pre-process all data that need to produce the explanation by using computer vision pipeline and make those data to textual information that easily adopt to current LLM. 
% To do this, we utilize prior research~\cite{chen2023iball} that includes datasets tracking the bounding boxes of both the ball and players. The dataset has the bounding boxes ball and players every frame. The ball owner is calculated by the closest player with ball. We utilize x and y position in court from the SportVU \cite{SportVU24} to classify tactics and recognize which player is in one-on-one relationship or defending to detect the actions.

\subsection{Tactic Detection}
\label{sec:tactic}
Identifying the tactics is crucial for answering tactic-based questions.
We classify a video clip's tactic by comparing five offensive players' movement patterns to those in a reference dataset \cite{tsai2017recognizing} that includes 134 annotated clips.
Each of these clips contains the temporal sequences of the five offensive players' coordinates and the associated offensive tactic, 
such as Back-Side Pick and Roll, Elevator, or Pin-Down. 
These patterns are represented by the five temporal sequence of their coordinates,
each denoted as $\{(x, y)\}_{i=0}^t$, with $x$ and $y$ are the on-court position, and $t$ marking the clip's end frame. 
% We classify the tactic of a given video clip by comparing its movement pattern with those in a reference dataset \cite{tsai2017recognizing} by finding the closest match.
Specifically,
we leverage the K-Nearest Neighbors (KNN) algorithm 
and identify the closest match in \autoref{fig:system}  (A-3).
The tactic type of the best match clip in this dataset is assigned to the current video clip in question.

Given the difference in sequence length between video clips,
it is impractical to use the traditional Euclidean distance for the KNN algorithm's distance metric. 
Consequently, we adopt FastDTW, a refined variant of Dynamic Time Warping (DTW), to overcome this challenge. 
FastDTW~\cite{salvador2007toward} can calculate the similarity distance between two temporal sequences of different lengths, 
thus offering a robust solution for our need.
\revision{We achieved an accuracy of 85.33\%. For more details, see the supplementary material, Section A.}

\subsection{Action Detection and Filtering}
\label{sec:action}
\begin{figure}[h]
\centering
\includegraphics[width=\columnwidth]{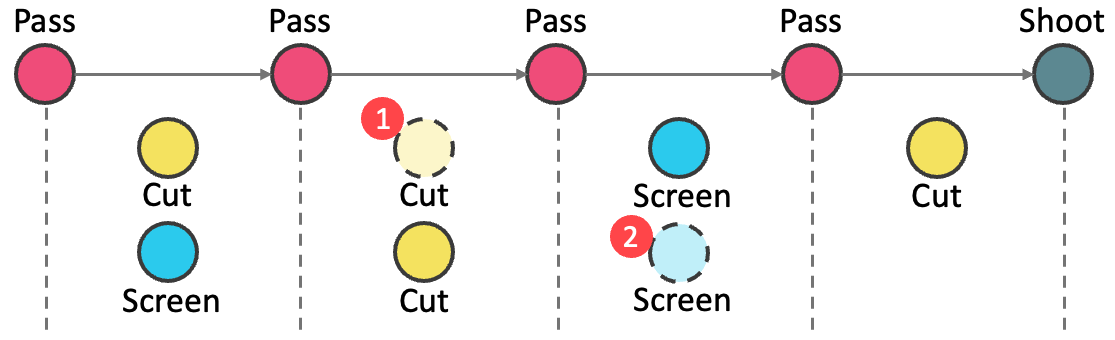}
\caption{An action list displays the series of actions performed by offensive players, including Pass, Cut, Screen, and Shoot. The primary actions are identified based on ball movement or movements that enhance scoring opportunities, such as Shoot or Pass the ball. To extract the primary actions related to Pass and Shoot, we set criteria to filter out secondary actions like Cuts and Screens, identified by actions 1 and 2 in red circles.}
\label{fig:action_graph}
\vspace{-0.4cm}
\end{figure}

\para{Action Detection.}
% Detecting actions from players' movements in the video plays an important role in explaining tactics and satisfying the requirement (R1). 
According to previous work~\cite{sicilia2019deephoops}, 
a tactic consists of a series of actions, each associated with a player.
Thus, besides the tactic type, we also need to detect the involved actions.
In this work, we focus on four most important actions in a basketball offensive tactic in \autoref{fig:system} (A-2).:

\begin{itemize}[leftmargin=*]

    \item \Pass{} occurs when a player transfers the ball to a teammate, who is in a more advantageous position to score. 
    A \Pass{} is detected by tracking \emph{ball ownership} changes  within the same team.
    
    \item \Cut{} is performed by a player who doesn't have the ball.
    The player moves swiftly from one court area to another to either create space or distance from a defender, thus enhancing offensive possibilities~\cite{courel2018inside, tian2019use}. 
    % This cut action is particularly effective against defenders as stationary players are easy to be guard.
    For detection, we segment the court into 10 sub-regions (e.g., key, post, wing), following a  taxonomy~\cite{unkown10glossary}.
    Then, a \Cut{} is detected if an offensive players moves at a speed of 6 feet per second or faster to a different sub-region~\cite{Wu2022OBTrackerVA}.

    \item \Screen{} refers to an offensive player's attempt to block or slow down a defender, thereby creating space and time for a teammate to move into a more advantageous offensive position or to take a shot~\cite{tian2019use}.
    From a technical standpoint, a screen occurs when an offensive player blocks a defender who is closely guarding another offensive player with possession of the ball. 
    Thus, to detect a \Screen{},
    % To identify when a screen is being set, 
    we analyze the \emph{distances} between players on the court. 
    A \Screen{} is detected if an offensive player changes their marking or covering player.
    
    \item \Shoot{} results in a change of possession regardless of whether it scores. 
    We detect a \Shoot{} action if the ball possession changes between the two teams.
\end{itemize}

The outputs of action detection are a list of actions,
each with its timestamp and the associated player.

\para{Action Filtering.}
Not all captured actions are relevant to the tactic of a team. Including redundant or non-crucial actions could diminish the user experience by cluttering the presentation with unnecessary details. 
Therefore, we propose to filtering the actions and keep the \emph{important} ones.
% After detecting the actions in the video clips,
% we still need to identify the \emph{important} ones.
A practical approach is to prioritize key actions integral to tactic implementation and scoring. 
According to Tian et al.~\cite{tian2019use}, 
\Pass{} plays a central role in executing tactics, 
while \Shoot{} is the end of a tactic.
Therefore, as illustrated in~\autoref{fig:action_graph}, 
we categorize \Pass{} and \Shoot{} as primary actions, 
while \Cut{} and \Screen{} as secondary actions. 
Then, we organize the actions chronologically and establish intervals between consecutive \Pass{}s (\autoref{fig:action_graph}).
% This distinction helps in filtering out irrelevant or less impactful actions that do not contribute meaningfully to the tactic explanation.

Next, we filter out the ineffective secondary actions,
including 1) \Cut{}s that are not followed by to a ball receive
and 2) \Screen{}s that are not positioned to benefit the ball handler or the intended receiver, based on the proximity of players and the location of the screen.
For example, 
if a player does not receive the ball after performing a \Cut{} (\autoref{fig:action_graph} 1), this action is considered as ineffective and removed from the interval.
Similarly, the \Screen{} in \autoref{fig:action_graph} (2) is excluded if it does not impact the pass, judged by the distance between where the screen is set and the locations of the ball pass before and after the screen.

After the filtering, we obtain a list of important actions that have direct impact to the outcome of the tactic. \revision{Our method achieved an F1 score of 73.93\% (Details n the supplementary material, Section A).}

\subsection{Retrieving External Statistics Data}
In addition to the data from the current video clip,
we also aim to provide the LLM with external meta data, such as the players basic information, team rankings, and historical performance.
To achieve this, \ourtool{} is equipped with a suite of tools designed for both in-game data analysis and external information retrieval, including 
programming tools like Pandas and external APIs like Google Search API, Wikipedia search API, and Statmuse API \cite{Statmuse24}.
The LLM can utilize these tools to extract necessary external statistics data by using a LLM framework named ReAct~\cite{yao2022react}, which enhances LLMs by enabling them to perform tasks and reason in a manner akin to human problem-solving.

\section{Narrative Agent}
\label{sec:llm_core}

To facilitate the understanding of the tactic (R2),
we aim to generate the explanation in a form of narratives (i.e., story).
This section introduce the design of the prompt to achieve this goal.
% We achieve this through \emph{prompt engineering}.

% The LLM achieve human-level performance across a wide range of tasks (e.g., professional benchmarks)~\cite{achiam2023gpt}, demonstrating its capabilities to comprehend the complex contexts of in-game decisions and generate reasonable answers to basketball strategic questions. (In our implementatoin, we used a OpenAI API with gpt-4-0125-preview model.) To produce storytelling-based strategic answer, the LLM is utilized to create narratives based on the user's question, a series of actions, tactical information, and prompts, as shown in \autoref{fig:system} (B-1). To be specific, the prompts play a crucial role in a consistent narrative and answer format to link visualization.

\para{A dynamic prompt with game Context.}
Rather than forwarding a user's query directly to the LLM, 
we enhance the input with extra game context from the current video clip. 
This includes \texttt{Player Information}, detected \texttt{Tactics}, and \texttt{Actions}, all sourced from the Data Processor (Sec.~\ref{sec:data_processor}).
This approach aligns with the Retrieval-Augmented Generation (RAG) framework, which has been proven effective in domain-specific tasks, enabling the LLM to draw upon a vast array of external knowledge without requiring retraining for specific applications.
For each query, the system dynamically loads game context relevant to the video clip and construct the prompt, guiding the LLM to tailor its responses to the clip in question.
Below is our prompt template, whose details can be found in the supplementary materials.

\begin{coloredquotation}
    \textbf{Prompt template:} 
    Please explain  \{ \texttt{user question} \} based on the following context:
    \begin{itemize}
        \item \{ \texttt{Player Information} \}
        \item \{ \texttt{Detected Tactics} \}
        \item \{ \texttt{Detected Actions} \}
    \end{itemize}
    
    Your explanation should follow the below constraints and formats:
    \begin{itemize}
        \item \{ \texttt{Constraints} \}
        \item \{ \texttt{Format} \}
    \end{itemize}
\end{coloredquotation}

% For every inquiry, our system dynamically incorporates relevant game context pertaining to the video clip, thereby allowing the LLM to generate responses that are specifically tailored to the clip in question. Further information on the format used for this structured data is available in the supplementary materials.
% Instead of directly sending the user's question to the LLM,
% we enrich the the user's input with additional game context of the current video clip,
% including \texttt{Player Information}, detected \texttt{Tactics}, and \texttt{Actions}, obtained from the Data Processor (Sec.~\ref{}).
% Such a practics follows the RAG framework~\cite{},
% which has demonstrated its effectiveness for domain-specific tasks~\cite{lewis2020retrieval, gao2023retrieval}, allowing the LLM to leverage a wide range of external knowledge without the need to be retrained for task-specific applications. 
% For each query, the system dynamically loads game context relevant to the video clip, enabling the LLM to tailor its responses to the specific clip in question.
% Details on the structural format can be found in the supplementary materials.

\para{Generating Explanations with Narratives.}
To enhance the comprehension of explanations, we adopt a top-down approach—presenting a tactic overview before delving into detailed actions. 
Our pilot experiments indicated that LLMs often struggle with generating comprehensive explanations in a single attempt. 
To address this, we've divided the task into two separate steps, in line with established best practices~\cite{shen2023data}. 

First, we prompt the LLM to produce an overview explanation of the tactics.
This overview succinctly summarizes the tactics, enabling fans to grasp the essentials of tactical unfolding.
Second, 
we add prompts to the LLM for generating a detailed, sequential breakdown of the tactics,
providing users with action-by-action insights into specific actions and decision-making processes. 
For each step, we incorporate different \texttt{constraints} to help the LLM generate different explanations aligned with each step's purpose.

% \begin{coloredquotationblue}
%     \textbf{Third person perspective narratives:}\\ 
%     \emph{``Draymond Green cuts from the top to the key aiming to create a scoring opportunity by disturbing the defense. Stephen Curry passes the ball to Draymond Green to create a better scoring opportunity.''}
%     % \begin{itemize}
%     %     \item \{ \texttt{Player Information} \}
%     %     \item \{ \texttt{Detected Tactics} \}
%     %     \item \{ \texttt{Detected Actions} \}
%     % \end{itemize}
%     \\\\
%     \textbf{First person perspective narratives:}\\ 
%     \texttt{Draymond Green}: \emph{``See that gap opening at the top of the key? I’m cutting there now.''}
%     \\
%     \texttt{Stephen Curry}: \emph{``Got it, This cut will really put pressure on their defense and open up the floor for us.''}
% \end{coloredquotationblue}

% “Draymond Green: See that gap opening at the top of the key? I’m cutting there now.
% Stephen Curry: Got it, This cut will really put pressure on their defense and open up the floor for us.”
%       “3rd_1”: “Draymond Green cuts from the top to the key aiming to create a scoring opportunity by disturbing the defense. Stephen Curry passes the ball to Draymond Green to create a better scoring opportunity. ”

% \noindent
This two-step approach not only refines the LLM’s output by incorporating detailed constraints but also safeguards against the generation of incorrect explanatory formats.

\begin{figure}[h]
\centering
\includegraphics[width=\columnwidth]{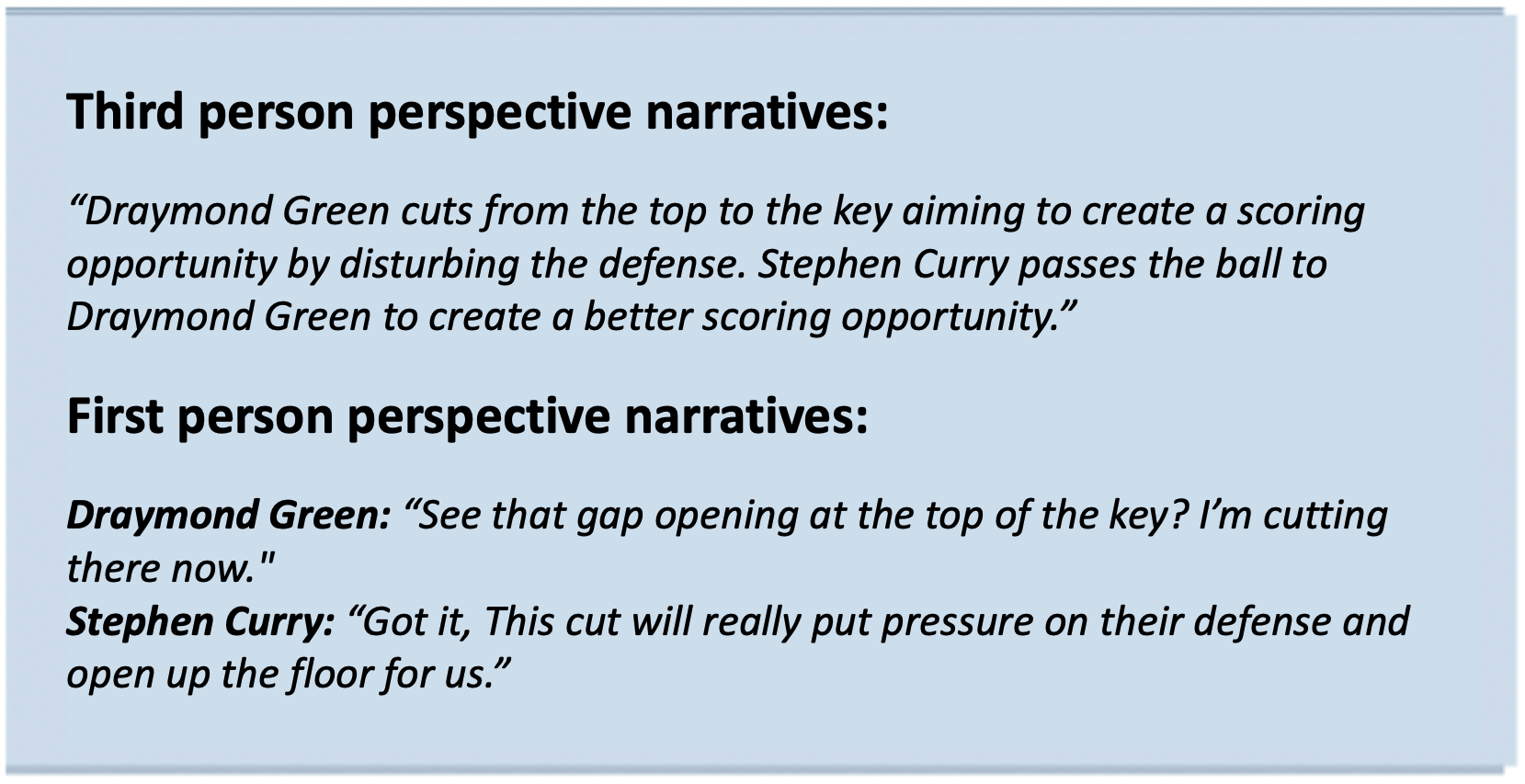}
\caption{The same tactics' explanation in various narrative perspectives.}
\label{fig:perspectivetext}
\vspace{-0.3cm}
\end{figure}

\para{Formatting Explanations with Various Perspectives.}
% The default narrative perspectives are third person.
The standard narrative perspective of LLM is third-person. 
However, to generate explanations from a first-person perspective—a task LLMs don't typically perform automatically—we introduce specific prompts that steer the model towards generating such responses. 
Specifically,
we prompt the model to frame its explanations as if part of a conversation or role-play dialogue between two players
(e.g., \textit{... answer should be a format of a conversation or a role-play dialogue between two players...}),
incorporating descriptions that evoke a first-person reaction to in-game actions (e.g., \textit{... screen elicit a surprised or shocked reaction from the other person...}).
An example format is provided to ensure answer consistency, given the challenge of maintaining uniform response formats due to the varied nature of actions and tactics. 
Such a prompt technique is applied to both the two aforementioned steps.
\autoref{fig:perspectivetext} shows an example of the explanation of the same tactics in different narrative perspectives.
Finally,
we also prompt the LLM to output the explanations in a 
structure format to facilitate their mapping to visualizations (Sec.~\ref{sec:vis_exp}).

\begin{figure}[ht!]
    \includegraphics[width=0.5\textwidth]{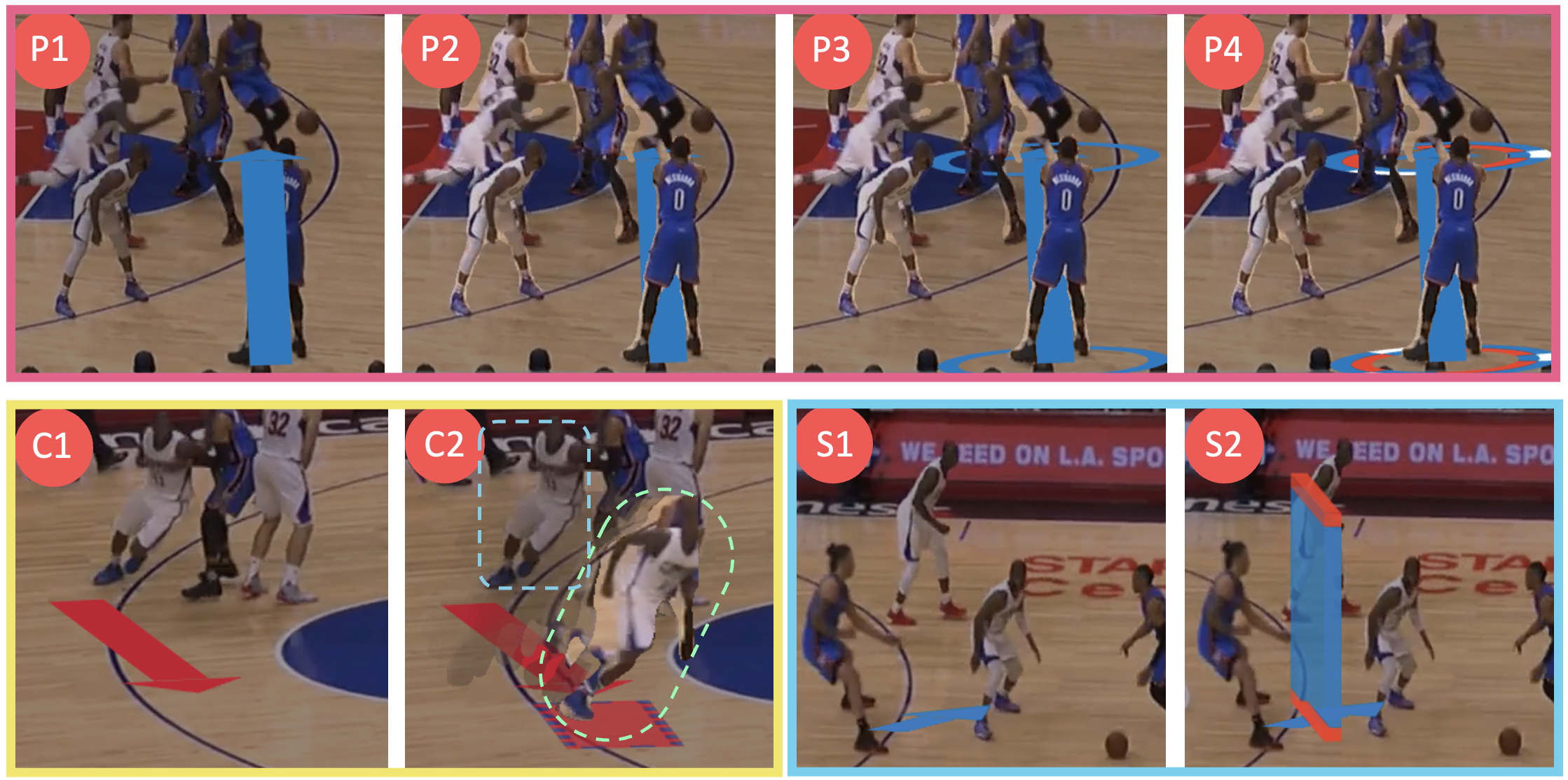}

    \caption{The iteration design process to design action visualizations (i.e., Pass, Cut, and Screen). From P1 to P4, we remove the occlusion and highlight the two players who send and receive the ball. For the cut, we indicate the exact location that a player will move with flash-forward animation from C1 to C2, while the screen demonstrate a wall to be easily identified a player set on screen from S1 to S2.}
    \label{figs:iter}
    \vspace{-0.5cm}
\end{figure}

\section{Visualizing Tactic Explanations}
\label{sec:vis_exp}
% Drawing on our design requirements in Sec. 3, we design a series of visualizations and user interfaces. 
% Our first requirement (interaction for complex questions, R1) does not necessarily require a response in the form of visualization, since the answer can be provided as simple textual information. 
% We focus on the second and third design requirements to address strategic and in-game decision-making questions (R2), and to facilitate storytelling from the different narrative perspectives (R3).

% Instead of presenting the explanations as pure text, 
% we propose to visually present the explanations (R3).
% As a tactic comprise a serious of actions,
% we decide to visualize the involved actions to present the explaination of
% the tactic.
% Moreover, we also need to visually present the narrative perspectives.
% Below, we introduce the design, respectively.

Rather than conveying explanations solely through text, 
we aim for visually representing them (R3). 
Since a tactic consists of a series of actions, 
we decide to visualize these actions to illustrate the explanation of the tactic. 
Additionally, it's essential to depict the narrative perspectives visually. In the following sections, we will detail the design approach for each of these visual designs.

\subsection{Visualizing Actions}

Given that the \Shoot{} action signifies the end of a possession, 
we focus on visualizing the other three actions: \Pass{}, \Cut{}, and \Screen{}. 
For clearer comprehension, we pause the video during these visualizations. The required data for rendering these visualizations, such as players' coordinates, bounding boxes, and specific frames, is obtained using computer vision models (\autoref{fig:system} A1).

\begin{figure*}
    \includegraphics[width=1.0\textwidth]{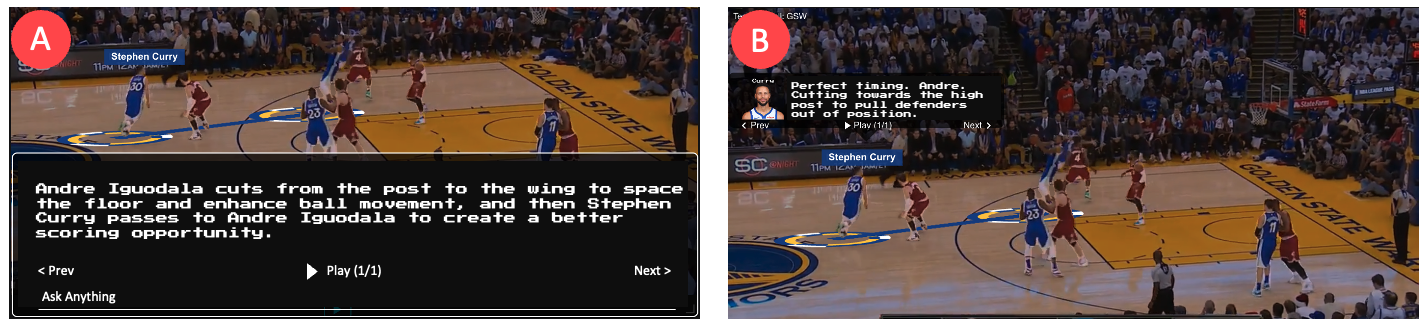}
    \caption{The figure (A) shows the third-person perspective narrative like commentaries, whereas the figure (B) demonstrates the first-person perspective by integrating the action visualizations and narratives around the players to make people more engaged and immersive.}
    \label{figs:perspective}
    \vspace{-0.5cm}
\end{figure*}

\para{Visualizing a \Pass{}.}
% The goal of a \textsc{Pass} in basketball is to transfer the ball to a teammate who is in a better position to score.
To visualize a \Pass{}, we aim to clearly delineate the dynamics of passing by highlighting the players involved in the pass—the sender and receiver—along with the ball's trajectory.
As demonstrated in~\autoref{figs:iter}(P4), 
our visual design employs two rotating circles that mark the sender and receiver, respectively, and an arrow beneath the players to indicate the ball's passing direction. %, thus avoiding occluding the scene.
Additionally, we incorporate a flash-forward effect~\cite{Chen2021Augmenting} to preview the players' subsequent movements.

The development of our visual design was an iterative process, detailed in~\autoref{figs:iter} 
 P1 to P4. 
Initially, we used a basic arrow to signify the change in ball possession and its direction (P1), but this approach proved problematic as it occludes the players.
To counter this, we repositioned the arrow beneath the players on the court (\autoref{figs:iter} P2).
However, the P2 design encountered visibility issues in crowded scenes.
Our subsequent iterations focused on enhancing visibility and understanding: 
we introduced circles beneath the players to emphasize their roles (P3) and further refined the design (P4) by animating the circles and arrow and adding a flash-forward effect for future movements, and employing team-specific colors for easy identification. 
This iterative approach, culminating in our final design, facilitates an intuitive and engaging visualization for fans to understand a \Pass{} action.

\para{Visualizing a \Cut{}.}
The purpose of a \Cut{} in basketball is for a player without the ball to move from one part of the court to another, aiming to create space from defenders and find better positions for shooting~\cite{courel2018inside, tian2019use}.
We visualize a \Cut{} by using an arrow to indicate the player's movement direction 
and an area visualization to demonstrate the specific court locations they are moving into (\autoref{figs:iter} C2).
Particularly, we incorporate a flash-forward effect,
which previews the player's trajectory before the actual movement occurs. 
In \autoref{figs:iter} (C2), a blue dashed line represents the player's position before the cut, and a green line indicates the position after the cut, effectively showcasing the movement path.
% A key feature of our cut visualization is the 
% inclusion of a , 

Our approach to visualizing a \Cut{} was also iteratively refined.
The initial design (\autoref{figs:iter} C1) depicted a \Cut{} with a simple arrow, highlighting the player’s direction but failing to convey the precise future location or the movement trajectory. 
To address these limitations, the refined design (\autoref{figs:iter} C2) integrates the arrow with area representation and the flash-forward effect.
This enhancement not only clarifies the direction and intent behind a player's \Cut{} but also provides viewers with a predictive insight into the player's positioning, 
enriching the overall understanding of the game’s dynamics.

\para{Visualizing a \Screen{}.}
A \Screen{} in basketball is executed to impede or slow a defender, thereby creating space and time for the offensive players. 
This allows the player with the ball to either move into a more advantageous position, dribble past opponents, or take a shot~\cite{tian2019use}. 
To visualize a \Screen{}, we aim to enable fans to easily identify when and by whom a screen is set against a defender
As shown in \autoref{figs:iter} S2, our design incorporates both an arrow and a depiction of a screen wall. 
This combination clearly distinguishes the involved players and the location where the screen occurs.

The initial design (\autoref{figs:iter} S1) utilized an arrow to indicate the screen action. 
However, this approach proved insufficient for clear identification, as it often blended with other actions and occluded by the presence of multiple players on the court. 
We thus improved it and concluded to the current design, which ensuring that fans can easily recognize and understand this crucial aspect of a \Screen{}

\subsection{Visualizing Narratives}
% In addition to visualize the actions,
Alongside action visualization,
we also need to present the narratives (i.e., the textual explanation of each action) together with their perspectives.
\revision{We initially intended to provide audio using text-to-speech technology. 
However, due to the time required to generate the speech, we decided to provide it in text form instead.}
\autoref{figs:perspective} demonstrates how the scenario of Stephen Curry passing to Andre Iguodala can be narrated from different viewpoints—illustrated through third-person (\autoref{figs:perspective} A) and first-person (\autoref{figs:perspective} B) perspectives.

The third-person perspective, akin to a commentator's overview (e.g., ``Andre Iguodala cuts from the post to the wing to space the floor...''), is traditionally used to explain basketball plays (\autoref{figs:perspective} A). 
This narration style fits seamlessly into visualizations using a single chatbox, streamlining integration without additional interface requirements.

Conversely, the first-person perspective, which captures players' emotions and dialogues (e.g., \emph{``Perfect timing, Andre. Cutting towards the high post to create a mismatch...''}), can confuse fans when presented in a traditional chatbox format. 
To address this, we adopt dialogue bubbles for first-person narratives, akin to comic book styles (\autoref{figs:perspective} B). 
These bubbles move with the players on screen, enhancing engagement by allowing users to interact with the narrative—navigating through the conversation with `previous', `play', and `next' controls.

% The third-person perspective is a common way to explain basketball plays, like a commentator's description (e.g., \textit{``Andre Iguodala cuts from the post to the wing to space the floor...''}), as shown in \autoref{figs:perspective} (A). 
% This perspective can easily be integrated into a visualization with one chatbox, which does not require any additional interfaces. 

% However, fans are confused and distracted with the traditional chatbox in the first-person perspective since the players express their emotions and thoughts, and call other team players by name (e.g., \textit{``Perfect timing, Andre. Cutting towards the high post to create a mismatch...''}). 
% We thus use an embedded dialouge buble to present 
% the first-person perspective narratives, similar to commics (\autoref{figs:perspective}B).
% The dialouge bubble will follow the players when they moves in the video, and provide interactions for the users to navigate the conversations by 
% clicking prev, play, and next.
% to integrate the narrative and visualization at the top or bottom of the players to indicate who is talking, as demonstrated in 
% a new user interface for 

% In \autoref{figs:perspective} (B), the integrated chatbox follows around a player while playing video and shows the interface that allows manipulation of players' conversations by clicking prev, play, and next.

\section{User Study}
We conduct a two-phased user study with basketball fans to evaluate the understanding, usability, and engagement of \ourtool{}.
Two experiments were conducted: a comparative study of narratives with first and third-person perspectives with and without visualizations, and an exploratory user study on the overall experience of \ourtool{}.

\subsection{Participants \& Experiment Set-Up}
We recruited 13 basketball fans (P1-P13; M = 13; Age: 23 - 33) via university mailing lists. Due to a technical issue, we were only able to collect P4's subjective feedback, excluding the task completion time and accuracy. Participants reported their fandom levels, including 3 novice, 2 casual fans, and 8 engaged fans. 
% We excluded die-hard fans or those already familiar with various strategies to focus on providing an understanding of game strategy for casual fans. 
In addition, participants reported their frequency of watching basketball in four different levels: 4 participants watched at least 1 game per week, 5 watched 2-4 games per month, 1 watched 11-23 games per year, and 3 watched 1-10 games per year. We selected two famous games: one between the Golden State Warriors and the Cleveland Cavaliers on December 25th, 2015, and the other between the Oklahoma City Thunder and the Los Angeles Clippers on December 21st, 2015. These games were featured as the best by the NBA in the 2015-16 season.
The two experiments were conducted in person on the same day, using a 14-inch laptop for setup. The study took about one hour to finish and all participants were compensated with a \$20 gift card.

\subsection{Study Design \& Measure}
\textbf{Introduction \& pre-survey (10 mins).} Before starting the user study, we introduced our user study and received the consent form.  We collected information about participants' backgrounds, such as fandom level and average watching frequency, through a pre-study survey. 

\para{Task 1: A comparative study (25 mins).}
Our first experiment investigates the understanding of strategies, the helpfulness of visualization and personified narratives. Participants first watched a game clip with an explanation, and were then asked to order a list of actions (e.g., pass, cut, and screen) to match the action sequence in the game based on their understanding. The explanation was delivered in one of the three conditions, including explaining strategies using pure text (\textit{Text}), text with visualization in third-person narrative (\textit{Third}), and text with visualization in first-person narrative (\textit{First}) conditions. The explanations were identical between \textit{Text} and \textit{Third} conditions. 
\revision{Explaining tactics often requires an in-depth understanding and detailed explanation, making the textual narrative necessary\cite{altavilla2014global, courel2017collective, tversky2002animation, choudhry2020once}. Therefore, the non-text version was excluded from the baseline.} 
We prepared twelve clips from both games evenly and tasked participants with their understanding of the strategy in the game. 
Each condition has four trials, including a practice and three actual trials, in which we collected task completion time and accuracy. 
The number and types of actions are evenly distributed among three conditions. 
In addition, the order of the conditions and the twelve clips were counterbalanced in the study. 

During the study, participants were introduced to three conditions and practiced the trials in the training session. In the study, participants watched the assigned game clip and used a drag-and-drop interface to move the order of the list of actions to match the sequence in the game. After completing all 12 trials, participants ranked their preferred explanation conditions and rated the helpfulness of the visualizations and narratives, providing subjective feedback in the questionnaire.

\para{Task 2: An exploratory study (25 mins).}
The second part of the user study involves a free exploration of \ourtool{} to compare participants' game-watching experience with tactic explanation videos on YouTube and using \ourtool{}. 
The participants chose either a first- or third-person perspective based on their preference, and analyzed a game selected from two games used in Task 1 (GSW vs. CLE or OKC vs. LAC) using \ourtool{}.
As a baseline for a typical tactic explanation approach for basketball games, 
we selected three tactic explanation videos from the popular basketball channels on YouTube\cite{genius, 5clever, offensive}, watching 2-3 minutes. 

In the study, the participants first watched a YouTube video randomly assigned from\cite{genius, 5clever, offensive}. After that, they were introduced to all features of \ourtool{} in a training video. Participants freely watched the game video and asked any questions at any time using \ourtool{}. 
\revision{To inform users when they can ask tactical questions, our system visually indicates which parts of video clips can be questioned and which cannot.}
Following previous work \cite{o2010development, lin2022quest, chen2023iball}, we measured user experiences using subjective rating questions in the post-study survey, including ``It was helpful'', ``It was fun'', ``I felt in control'', ``I felt encouraged'', ``I am likely'', and ``I felt engaged'' when using \ourtool{}, and collected feedback from the survey and think-aloud methods during the study.

\section{Study Results}

We present the results of two tasks in our user study 
and discuss how visualizations and different narrative perspectives affect the understanding of game strategy. Besides, we discuss how \ourtool{} differs from existing strategy explanation videos in enhancing game understanding.

\subsection{How Do Three Different Conditions Affect Understanding Strategy Explanations?}
In Task 1, participants matched the sequence of actions in twelve clips under three conditions: 
1) pure text in a third-person perspective (\textit{Text}), 
2) text with visualization in a third-person perspective (\textit{Third}), 
and 3) text with visualization in a first-person perspective (\textit{First}).
We presented the findings on how visualization and personified narrations affect strategy comprehension, task performance, and overall experiences.

\subsubsection{Watching Time Increased Without Affecting Accuracy}
We measured the accuracy of nine game clips in the trials from 12 participants, excluding three practice clips. The accuracy of matching the actions to game strategy for the \textit{Text} and \textit{Third} conditions were identical at 72.22\% (26 out of 36), while the \textit{First} condition was slightly lower at 69.44\% (25 out of 36). The average watching time for the \textit{First} condition (86.27s) exceeded that of the third-person perspective (\textit{Text}: 69.44s and \textit{Third}: 68.72s). However, the average solving time, which measures the time participants took to complete ordering actions in the trial after watching the clip, was similar across all three conditions (\textit{Text}: 40.33s, \textit{Third}: 39.94s, and \textit{First}: 44.53s). After conducting a normality test, we found that none of the conditions followed a normal distribution, so we performed a Kruskal-Wallis test. This test revealed a significant difference in watching time among the three conditions ($p$= 0.02 and $H$= 7.765) but not in solving time ($p$= 0.89 and $H$=0.24). Dunn’s post-hoc test further identified a significant difference between the \textit{Text} and \textit{First} conditions ($p$=0.046) in watching time.

Overall, the results show that the \textit{Third} condition has a similar watching time to the \textit{Text} condition, with visualizations aiding users in understanding the main actions in the clip. 
Participants found visualizations in \textit{Third} helpful, as noted \textit{``The explanations with visualizations help more than the text in general (P9)''}. 
In addition, participants spent 25\% more time watching the video with first-person narrative in \textit{First} than the third-person narrative in \textit{Text} and \textit{Third} condition.
This is likely due to that the \textit{First} condition involves more interactions than the other two conditions due to the personified narratives presented between a pair of players, leading to longer time to interact and engage in the video.
Participants did not find this interaction impeding their understanding, but instead felt that \textit{First person perspective text was more enjoyable and simpler to understand than the third person perspective text (P13)}. \textit{First} condition was found to be most helpful and enhanced their confidence, as shown in Sec.~\ref{sec:enhance}.

% it's fun and easier to follow than 3rd text.
% Users must click the "Next" button to navigate through conversations, resulting in longer game-watching times.

\begin{figure}[t!]
    \includegraphics[width=0.5\textwidth]{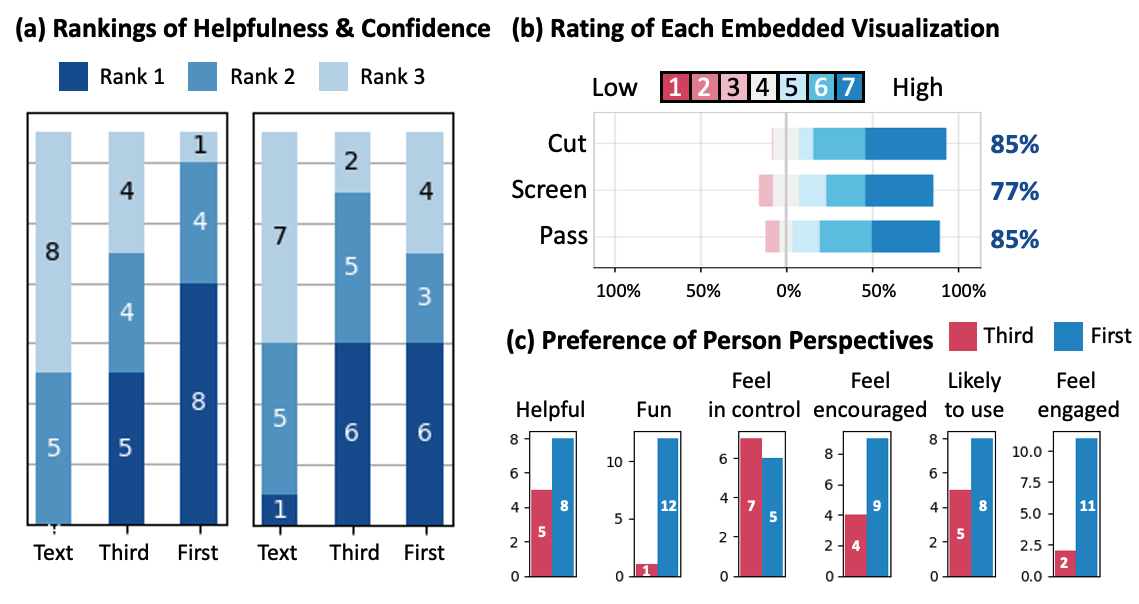}
    \caption{A task 1 user study results. Figure (a) reveals that the First person perspective ranks highest in helpfulness and confidence across three conditions. Figure (b) indicates positive participant ratings for each embedded visualization's helpfulness. Figure (c) compares usability across two different narrative perspectives.}
    \label{figs:task1}
    \vspace{-0.5cm}
\end{figure}

\subsubsection{First-Person Narrative Enhanced Feeling of Helpfulness and Confidence}
\label{sec:enhance}
\autoref{figs:task1} (a) reveals that the \textit{First} condition was rated highest in helpfulness and enhancing their confidence in understanding by 8 and 6 participants out of 13 participants, respectively, followed by the \textit{Third} condition with 5 and 6 participants, and the \textit{Text} condition with none and 1 participant. 
Participants highlighted the advantages and drawbacks of the \textit{First} and \textit{Third} conditions. 
Six participants noted that placing the chat box near players was helpful and not distracting, stating that \textit{''First person perspective chat box comes closer to players, which helps me better understand tactics. (P1)''}.
They also appreciated the interactive ``Play'' button for navigating conversations, feeling an experience similar to playing a video game, with one saying, \textit{``I am `in control' of the players, almost like a video game.''}
However, some found the \textit{First} condition confusing due to its abundant text and lack of resemblance to real commentary, with feedback like, \textit{``First condition felt a bit informal and not realistic, (P12)''}.

Four participants preferred the \textit{Third} condition for its realism and detailed information, commenting, \textit{``The Third condition was more effective in conveying detailed information.''}
However, they reported losing focus due to the need to switch attention between visualizations and text, with one mentioning, \textit{``I was distracted by moving my eyes between players and chat box (P1)''}.
% cons
% pros third
% cons

Overall, participants found the \textit{First} and \textit{Third} narrative perspectives more helpful than \textit{Text} alone in enhancing their understanding of game strategies. With these visual narratives, participants felt more confident in grasping game strategies, with participants equally favoring the \textit{First} and \textit{Third} perspectives. 
The \textit{First} offered interactivity and embedded placement of the visualizations, while the Third provided realism and clearer comprehension.

\subsubsection{Embedded Visualizations Help Identifying Player Interactions and Ball Movements}
As shown in the \autoref{figs:task1} (b), 
% the participants rated the helpfulness of three embedded visualizations on a scale from 1 (strongly disagree) to 7 (strongly agree). 
all the three visualizations received positive ratings from the majority, including Cut (85\%), Screen (77\%), and Pass (85\%), confirming the usefulness of the three visualizations for understanding strategic explanation.

Participants, especially novice and casual fans, commented that these action-specific visualizations significantly aided their understanding of basketball plays.
In particular, visualizations significantly aided in identifying ``Screen'' and ``Cut'' actions, allowing participants to comprehend complex plays like off-ball movements and interactions between offensive and defensive players more easily.
This result aligns with our main goal of elucidating off-ball movements by pinpointing the screen's location and direction with the Screen visualization and adding a flash-forward effect to demonstrate the next player move in the Cut visualization.
Furthermore, participants found the Pass visualization very helpful for its clear depiction of ball movement directions, senders, and receivers. 
For instance, P7 mentioned, \textit{``I sometimes miss who receives the pass, but with visualization, I could easily notice it.''} 
This feedback confirms the primary design goal of the Pass visualization in highlighting the players initiating and receiving passes. 

However, watching the simultaneous presentation of the Screen visualization and text presents a challenge, as noted by P11: \textit{``It would take some time to get used to seeing the text and visualization, since it shows a lot at once.''} 
This indicates a learning curve for users when simultaneously comprehending the visualization and text integration.

% P11: However, it definitely would take some getting used to as the text + the visualizations can be a lot one the screen at once.

% P7: As a casual basketball fan, it was harder for me to identify "screen" or "cut" actions without the visualization.
% P7: For the "pass" I could easily identify without the visualization. But the visualization helped me understand who the receiver is.
% P12: I think the pass is pretty obvious for anyone watching where the ball is going, but the screen visualization was helpful, since that is more of an off-ball movement. The cut visualization also helps off-ball understanding.
% P13: The pass is more noticeable, but the screen is not clear when watching the game.

\subsubsection{First-Person Enhances Fun and Enjoyment, While Third-Person Allows Feeling of Control and Formality}
\autoref{figs:task1} (c) shows the usability ratings for the \textit{First} and \textit{Third} person perspective narrative designs.
This result revealed a clear preference for \textit{First} person perspective narrations over \textit{Third} person on fun and engagement by 12 and 11 participants out of 13. 
This preference is attributed to the enjoyment of story-based narratives and a user interface design that simulates players engaging in conversation with one another.
P2 praised the narrative for its use of nicknames, expressing \textit{``I like the story-based narrative where they even called each other nicknames like Matty''}. 
The \textit{First} person perspective was also seen as creative and interactive, with P11 finding the player communication innovative and enjoying the conversational navigation. 
P11 noted that \textit{``I felt most interactive and creative when manipulating players' conversations.''}
Moreover, participants found this perspective helpful in understanding player thoughts and intentions, enhancing their grasp of overall tactics, as P5 mentioned \textit{``It is easier to understand each player's intention...''}
However, P8, a former university team player, criticized the dialogue in this condition for lacking realism and not accurately reflecting players' in-game thought processes.
As a result, our results reveal that the first-person perspective enhances engagement, encouragement, and fun, supporting the findings in previous research~\cite{mulcahy2016positioning, chen2021changing}.

On the other hand, \autoref{figs:task1} (c) demonstrates that from the post-survey feedback, 7 and 5 participants found the \textit{Third} person perspective narratives make them feel in control and found it more helpful, respectively. 
The text in third-person perspective is viewed as more concise and accurate for strategy comprehension compared to the first-person perspective. 
P3 stated \textit{``Third person perspective avoids unnecessary information like `Now, it is time to shoot'.''}. 
% This method effectively outlines strategic action sequences without unnecessary details.
Furthermore, some participants favored the third-person perspective for its conventional and formal explanatory style like game commentaries, mirroring their typical basketball viewing experiences.
As P6 mentioned, \textit{``Third person perspective seems more standard, formal, and familiar to me.''} 
Additionally, P9 noted that the third-person perspective is easier to follow and control with its predictable layout, highlighting that \textit{``It's easy to lose attention while following the chatbox above the players, and the third-person perspective is easier to control with the static chatbox navigation button.''}
According to other research~\cite{miall2001shifting}, shifting the perspective from third-person to first-person requires additional reading time and effort, causing discomfort for users who are accustomed to the third-person view.

In sum, participants' feedback distinctly indicates that first-person narratives provide a greater sense of fun and engagement, inspiring users to explore further, whereas third-person narratives, with their familiar format, enhance the feeling of control.

\begin{figure}[t!]
    \includegraphics[width=0.5\textwidth]{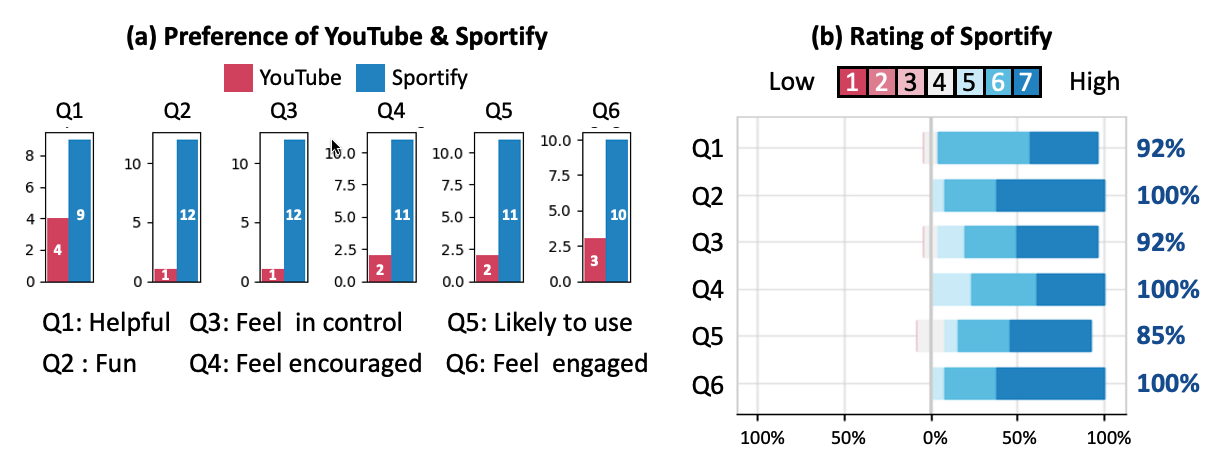}
    \caption{\ourtool \space received better results compared to existing tactical explanation videos, with positive outcomes in all questionnaire of usability.}
    \label{figs:task2}
    \vspace{-0.5cm}
\end{figure}

\subsection{How Does \ourtool{} Compare to Traditional Basketball Strategy Explanation Videos?}
Task 2 evaluated \ourtool{}'s usability by asking
participants to watch an existing strategy explanation video and then freely used \ourtool{}. 
% We compared the usability of these two methods.

% \subsubsection{\ourtool{}: infinite contents and understandable strategy explanation}

Participants predominantly favored \ourtool{} over the existing strategy explanation videos, as shown in numbers of participants on metrics of helpful (9), fun (12), feel in control (12), feel encouraged (11), likely to use (11), and feel engaged (10) in \autoref{figs:task2} (a).
The majority of participants valued \ourtool{}'s capability to generate personalized answers to their queries and provide infinite content.
Notably, two casual fans mentioned that while the existing videos left their questions unanswered, \ourtool{} provided the desired information on demand. 
P5 highlighted \textit{``\ourtool{}'s advantage of offering limitless, real-time content compared to the fixed material in traditional videos.''}
The tool's explicit presentation of strategies and motivations was especially helpful, with one user noting, \textit{``It helped me see the clear motivation and strategy explanation from a first-person view (P4)''}, underscoring \ourtool{}'s effectiveness in delivering detailed play insights.

On the other hand, some participants preferred the existing explanation videos for their structured content and insights from professional analysts.
Engaged fans with an understanding of basic strategies appreciated the depth of professional analysis, as P4 noted, \textit{``More rigorous analysis of plays with explanations from professional analysts.''} 
Additionally, the advocates of the video highlighted their ease of use, pointing out that they demand less effort to navigate, with P9 stating, \textit{``The video is easier to watch since it requires less effort.''}

\autoref{figs:task2} (b) illustrates \ourtool{}'s usability ratings on a scale from 1 (strongly disagree) to 7 (strongly agree), with all aspects receiving positive feedback (above 85\%). 
\ourtool{} received notable 100\% positive responses in fun, encouragement, and engagement. 
Users particularly enjoyed the first-person perspective for its dynamic explanation of tactics and player thought processes, as P4 highlighted: \textit{``It described the thought processes of players in a fun way, which helped me understand.''} 
The ability to ask personalized questions at their own pace encouraged participants to explore \ourtool{}, exploring specific areas of interest and seeking detailed information. 
P11 describes \textit{``It moves at your own pace, allowing you to probe the system with more specific questions about any play.''} 
The enhanced accessibility and in-depth analysis motivated users to engage more deeply with the strategies, with P12 appreciating the \textit{``in-depth analysis that focuses on overall strategy, especially showing the thought processes and reasoning among players.''}
P5 commented a strong positive feedback in terms of ``likely to use'', citing its real-time capabilities as particularly appealing: \textit{``If this system is public, I will absolutely try it. 
The real-time ability is really cool.''} 
In summary, eleven participants found the system extremely helpful, praising its accurate and sensible explanations and the freedom to inquire at will.

\section{Discussions}

% Implications of VQA Systems for Sports Videos}
We summarize the design insights derived from developing our system, 
along with feedback and observations from the user studies.

% audio
% concise options
% \subsubsection{Basketball narratives person perspective: different preferences from the general stories}
\subsection{Sports Narratives: Various Preferences for Perspective}

% Initially, we believed that the choice between first and third-person perspectives might depend on the viewers' backgrounds, such as their level of fandom, knowledge, and frequency of watching basketball. 
Previous studies~\cite{mulcahy2016positioning, green2004transportation} have shown that readers familiar with the events and emotions of a story's protagonist are more likely to choose a first-person perspective, while those with less experiences prefer a third-person perspective. 
Based on this, we thought that novice or casual fans less familiar with basketball would choose the third-person perspective, while engaged fans would select the first-person perspective.

However, during free exploration of our system, participants showed an equal preference for both perspectives: seven for first-person and six for third-person. 
Among the participants, two novices and one casual fan chose the first-person perspective, as did four out of eight engaged fans. This trend was consistent across various levels of basketball knowledge and viewing frequency.
Unlike general storytelling, a crucial factor here is the need to understand complex basketball strategies. As mentioned in~\cite{chen2023iball}, novice and casual fans tend to enjoy games more for their entertainment value and likely prefer the first-person perspective for its immersive experience. In contrast, engaged fans focus on technical aspects and strategies, often seeking a deeper understanding, and might lean towards the third-person perspective for its broader overview.
Therefore, this mixed conclusion likely stems from individual familiarity with the protagonist and the need for a deeper understanding of the narratives.

\subsection{Using Text-Based LLMs Instead of Multi-Modal LLMs}
In developing \ourtool{}, we experimented with utilizing current state-of-the-art multi-modal LLMs~\cite{chen2023llava, zhang2023gpt4roi, liu2024visual} that are capable of processing images or videos to interpret visual content. 
However, we found that applying multi-modal LLMs is still challenging in the sports domain, which is characterized by numerous players and complex dynamics and interactions.
Instead, we transformed visual data from the videos into textual information using a computer vision pipeline, providing the LLM with detailed information such as actions and tactical insights to facilitate answer generation. In this framework, the LLM served as a central hub or ``brain'' to synthesize all information and make inferences.
In the future, we anticipate that advancements in multi-modal LLMs' capabilities in understanding images and videos will enable a more integrated approach to system development, offering a seamless, all-encompassing solution. 
One limitation of our study was the inability to provide defensive tactics due to the lack of data, but we believe that this will be possible with multi-modal LLMs in the future.

% However, for now, we suggest to generate the desired information through this approach.
% other methods (KNN) as tools
% LLM try to merge all information and finish tasks
% This paradigm is different all in one modal.

\subsection{Presenting LLM-Generated Text in Videos}
Explaining the dynamics of sports is challenging. 
Previous research has mainly focused on enhancing understanding through embedded visualizations, 
often neglecting textual descriptions. 
With text descriptions generated by LLMs, we have explored different methods to 
present these descriptions in the video.

% with embedded visualizations in the video.
% Our research combined a storytelling approach with embedded visualization. 
% Thus, many sports videos support their explanations with commentary to elucidate these dynamics and situations. 
% Previous papers have explored only embedded visualization without including narrative elements like commentary. 

% 
\para{\revision{Presenting Text with Visualizations.}}
\revision{
% Given the advanced capabilities of LLMs, 
% Many studies and products aim to combine LLM-generated text with visualizations. 
Presenting LLM-generated text with visualizations in the video is an intuitive method.
However, 
simply integrating LLM-generated text and visualizations without careful design consideration could hinder the user experience. 
As identified in our paper, placing third-person perspective text with visualizations forces viewers to switch between action visualizations around players and text in the chat box, 
leading to a loss of concentration and degrading the user experience. 
This highlights the need for careful visualization designs and user interfaces.
While LLMs can potentially identify positions to place text with visualizations automatically, they are not specifically designed for particular tasks, making it challenging to achieve complex output. 
Therefore, simplifying the problem through an intermediary step and then subsequent stages such as visualization helps us achieve better results~\cite{sultanum2023datatales, shen2023data}.
% an intermediate transformation or linking process is necessary
% Therefore, 
}

% Given the advanced capabilities of LLMs, many studies and products will combine LLMs with visualization.
% However, simply integrating LLM and visualization together without careful design consideration could hinder the user experience.
% As identified in our paper, placing third-person perspective text with visualization forces readers to switch between action visualizations around players and text in the chat box, leading to loss of concentration and degrading of the user experience. This points to the need for careful visualization designs and user interfaces.
% Furthermore, since directly translating narratives into visualizations is difficult, an intermediate transformation or linking process is necessary \cite{sultanum2023datatales, shen2023data, jiang2023graphologue}. 
% LLMs are not models specifically designed for particular tasks, sometimes making it challenging to achieve the complex output. Therefore, simplifying the problem through an intermediary step and then subsequent stages such as visualization help us gain better results.

% We investigated how visualizations should be designed and their effect on user understanding.
% We implemented this through "actions". 
% [Chunggi, add a couple sentence to briefly describe why this step is useful and what other people should learn in the future]

\para{\revision{Presenting Text using Audio.}} 
\revision{Another alternative we considered was the use of audio. We attempted to convert the generated text into audio and play it together with the visualizations, but found that this process was time-consuming. Consequently, we decided to omit this feature as it could degrade the user experience. However, during our user surveys, two participants inquired about the availability of audio. We believe that incorporating audio with visualizations could potentially allow for a more seamless integration.}

% \subsubsection{LLM Agent - fundamental system LLM as brain}
% take away and insights
% try to find RW - other people insights and compare them, or saying no one has done before
% What is the next

% X narration from previous works -> how impact design visualization and users' understanding

% Stroytelling overall + embedded visualization
% How narrative impact users' understanding in dynamic situations
% Adding to narrative components
% how this contribution of narratives to embedded visualization

% can explain complex actions
% generalize to other domains

% data movie

\subsection{Designing User-Centered LLMs for Domain-Specific Tasks}
\para{\revision{Strategies for Ensuring Consistent LLM Responses.}} 
\revision{Given the unfamiliarity of first-person narratives to LLMs, ensuring a uniform style of response presented a significant challenge. The complexity increased when attempting to generate conversation-like formats where multiple players interact and address each other by name. To achieve consistency in responses, we applied multiple strategies:}

\revision{
First, we utilized a template-based approach within the prompts, integrating various constraints as suggested by previous studies~\cite{shen2023data}. Despite these efforts, the LLM still struggled to deduce the logic or reasoning behind in-game decisions from textual input alone. To aid in this reasoning process, we explicitly outlined the rationale for 3 to 5 actions within the prompt.
}

\revision{
Additionally, we employed ReAct (Reason + Act)~\cite{yao2022react}, which generates human-like task-solving trajectories and prevents error propagation. We believe that these approaches help prevent issues of hallucination. However, if hallucinations still occurred despite these methods, users were encouraged to rephrase their questions.
}

% To enhance the accuracy of explanations, we conducted an experiment with prompt engineering. 
% Given the unfamiliarity of first-person narratives to the LLMs, ensuring a uniform style of response presented a significant challenge. The complexity further increased when attempting to generate conversation-like formats where multiple players interact and address each other by name. To achieve consistency in responses, we utilized a template-based approach within the prompts, integrating various constraints as suggested by previous studies~\cite{shen2023data}.
% Despite these efforts, the LLM still struggled to deduce the logic or reasoning behind in-game decisions from textual input alone. To aid in this reasoning process, we explicitly outlined the rationale for 3 to 5 actions within the prompt. 
% \revision{In addition, we utilized ReAct (Reason + Act)~\cite{yao2022react}, generating human-like task-solving trajectories and preventing error propagation. We believe that these approaches prevent issues of hallucination. However, if hallucinations still cannot be prevented by the above two methods, the user asked a different question in another way.}

\para{\revision{Selecting Key Information for Enhanced LLM Performance.}}
\revision{
The number of actions detected in a video is usually too many for the LLM to reason through effectively. 
By filtering these actions to include only essential information for the LLM, we streamlined the data, akin to applying an importance score for selecting critical information~\cite{park2023generative}.
% Therefore, we apply an importance score to select critical actions~\cite{park2023generative}, providing only essential information to the LLM. 
These strategies resulted in a more precise system, surprising users with its accuracy in providing explanations during the user study. 
Feedback from the study highlights the critical role of fusing domain-specific knowledge to curate information for the LLM in developing LLM applications. 
Our approach of guiding the LLM with key information proved to be effective and satisfying.
}

% Initially, the plethora of actions was too vast to visualize simultaneously. 
% By filtering these actions to include only essential information for the LLM, we streamlined the data, akin to applying an importance score for selecting critical information~\cite{park2023generative}. 
% These strategies led to the creation of a more precise system, surprising users with its accuracy in providing explanations during the user study. 
% Many participants were amazed by the system's capability to offer in-depth rationales.
% This outcome highlights the critical role of adopting a user-centric methodology in the development of LLM application. Our approach of guiding the LLM with examples and providing main information to enhance its performance was shown to be effective and satisfying.

% \para{\revision{Various User Needs.}} 
% We also noted that not every fan seeks intricate tactical insights or understands the players' thought processes. A number of participants favored succinct explanations, indicating a user preference for customizable narrative detail levels. This suggests that offering users the ability to select their desired level of detail could significantly enhance their experience.

\para{\revision{Addressing Various User Needs.}}
We observed that not every fan seeks intricate tactical insights or understands the players' thought processes. Several participants favored succinct explanations, indicating a preference for customizable narrative detail levels. This suggests that allowing users to select their desired level of detail could significantly enhance their experience.

\subsection{Limitations}
\revision{
% Another embedded visualization
% action name
% sample size
The sample size of our study is comparable to other similar sports visualization papers~\cite{lin2022quest, chen2023iball}. 
Besides, the focus of our study is on qualitative feedback rather than quantitative results. Through qualitative feedback, we found that not only novices but also engaged fans benefited from and were immersed in understanding the tactics. 
Nonetheless, further experiments with a larger user base are beneficial.
% Our study used a smaller sample size compared to other similar sports visualization papers . However, the our main focus was on qualitative feedback rather than quantitative results. 
%We found that not only novices but also engaged fans benefited from and were immersed in understanding the tactics from the qualitative results.
% We found that our system was helpful not only for beginners but also for engaged fans, and we believe 
In real-world settings, our VQA system faces challenges related to data precision, processing speed, and user diversity. The precision of 3D tracking data from single monocular videos is often insufficient, but this can be improved with 3D vision models using multiple-angle videos or extra sensors. The system's multiple pipelines require significant processing time, hindering real-time performance, though this can be mitigated with better machinery and lighter models. Additionally, customizing conversation content to suit diverse user needs and skill levels, including adjustable explanation detail and length, is crucial.
Lastly, our initial focus was on explaining tactics through key actions in the proposed visualizations. However, more diverse visual explanations are needed to fully support the tool's practical use.
% For real-world settings, there are challenges related to data and processing speed. 
% Our VQA system operates based on pre-tracked data, requiring real-time trajectory information of players and players’ statistical data. 
% To process complex and diverse data, the system is expected to take considerable time to process information through multiple pipelines.
% We believe these issues can be effectively addressed with the advancement of various sensors and technology for the real-world deployment. 
% Lastly, additional visualizations (e.g., highlighting players, ball trajectory, etc) can be integrated in the future and be helpful for users to understand not only tactics but also entire games.
}
\section{Conclusion}
Our work introduces \ourtool{}, an innovative VQA system. 
It significantly enriches the basketball watching experience by integrating embedded visualization with personalized narrative explanations of tactics. This novel system enables fans to investigate understanding the game through both statistical queries and complex tactical questions. By employing a computer vision pipeline and leveraging a LLM to generate insightful explanations of players' logic and reasoning, \ourtool{} transforms the paradigm from passive watching to an interactive, engaging exploration of basketball. Our user studies evaluate three different conditions and two narrative perspectives in a comparative study, as well as the usability of our tool in a free exploratory study. The results reveal that \ourtool{} significantly enhances users' comprehension of tactics made during games and elevates user engagement and experience beyond what is offered by existing tactic explanation videos. Moreover, narration from a third-person perspective aids in providing detailed explanations of the game, while first-person perspective increases fans' enjoyment and engagement with the game.

\section*{ACKNOWLEDGMENTS}
This work is supported by SEAS Graduate Fellowship, NSF grant IIS-1901030, NIH grant R01HD104969,  NSF grant III-2107328.

\newpage

\bibliographystyle{abbrv-doi-hyperref}

\bibliography{template}

\begin{thebibliography}{10}

\bibitem{nbaviewership}
23 amazing nba viewership statistics in 2024.
\newblock \url{"https://playtoday.co/blog/stats/nba-viewership-statistics/"}.
\newblock Accessed on March 25, 2024.

\bibitem{5clever}
5 clever nba set plays and strategies explained.
\newblock \url{"https://www.youtube.com/watch?v=Fd3MzuHKHHI"}.
\newblock Accessed on March 21, 2024.

\bibitem{genius}
6 genius nba plays explained.
\newblock \url{"https://www.youtube.com/watch?v=lpR9Fp84XPw&t=146s"}.
\newblock Accessed on March 21, 2024.

\bibitem{courtvision}
Court vision.
\newblock \url{"https://www.clipperscourtvision.com/"}.
\newblock Accessed on March 25, 2024.

\bibitem{popularSports}
The most popular sports in the world.
\newblock \url{"https://www.worldatlas.com/articles/what-are-the-most-popular-sports-in-the-world.html"}.
\newblock Accessed on March 25, 2024.

\bibitem{SportVU24}
Nba sportvu dataset.
\newblock \url{"https://paperswithcode.com/dataset/nba-sportvu"}.
\newblock Accessed on March 21, 2024.

\bibitem{nbawebsite}
Nba website.
\newblock \url{"https://www.nba.com/"}.
\newblock Accessed on March 25, 2024.

\bibitem{offensive}
One of my favorite nba offensive concepts.
\newblock \url{"https://www.youtube.com/watch?v=_wA4Fpzx08s"}.
\newblock Accessed on March 21, 2024.

\bibitem{secondspectrum}
Second spectrum.
\newblock \url{"https://www.secondspectrum.com/"}.
\newblock Accessed on March 25, 2024.

\bibitem{sportvucamera}
Sportvu camera system in nba.
\newblock \url{"https://www.statsperform.com/team-performance/basketball/optical-tracking/"}.
\newblock Accessed on March 25, 2024.

\bibitem{Statmuse24}
Statmuse.
\newblock \url{"https://www.statmuse.com/"}.
\newblock Accessed on March 14, 2024.

\bibitem{vizlibero}
Viz libero.
\newblock \url{"https://www.vizrt.com/products/viz-libero."}.
\newblock Accessed on March 25, 2024.

\bibitem{altavilla2014global}
G.~Altavilla, G.~Raiola, et~al.
\newblock Global vision to understand the game situations in modern basketball.
\newblock {\em Journal of Physical Education and Sport}, 14:493--496, 2014.

\bibitem{chen2021changing}
M.~Chen and R.~Bunescu.
\newblock Changing the narrative perspective: From deictic to anaphoric point of view.
\newblock {\em Information Processing \& Management}, 58(4):102559, 2021.

\bibitem{chen2023llava}
W.-G. Chen, I.~Spiridonova, J.~Yang, J.~Gao, and C.~Li.
\newblock Llava-interactive: An all-in-one demo for image chat, segmentation, generation and editing.
\newblock {\em arXiv preprint arXiv:2311.00571}, 2023.

\bibitem{choudhry2020once}
A.~Choudhry, M.~Sharma, P.~Chundury, T.~Kapler, D.~W. Gray, N.~Ramakrishnan, and N.~Elmqvist.
\newblock Once upon a time in visualization: Understanding the use of textual narratives for causality.
\newblock {\em IEEE Transactions on Visualization and Computer Graphics}, 27(2):1332--1342, 2020.

\bibitem{Chu2021TIVEEVE}
X.~Chu, X.~Xie, S.~Ye, H.~Lu, H.~Xiao, Z.~Yuan, C.~Zhu-Tian, H.~Zhang, and Y.~Wu.
\newblock {TIVEE: Visual Exploration and Explanation of Badminton Tactics in Immersive Visualizations}.
\newblock {\em IEEE Transactions on Visualization and Computer Graphics}, PP:1--1, 2021.

\bibitem{courel2018inside}
J.~Courel-Ib{\'a}{\~n}ez, A.~P. McRobert, E.~Ortega~Toro, and D.~C{\'a}rdenas~V{\'e}lez.
\newblock Inside game effectiveness in nba basketball: Analysis of collective interactions.
\newblock {\em Kinesiology}, 50(2.):218--227, 2018.

\bibitem{courel2017collective}
J.~Courel-Ib{\'a}{\~n}ez, A.~P. McRobert, E.~O. Toro, and D.~C. V{\'e}lez.
\newblock Collective behaviour in basketball: a systematic review.
\newblock {\em International Journal of Performance Analysis in Sport}, 17(1-2):44--64, 2017.

\bibitem{de2023visual}
A.~C. A.~M. de~Faria, F.~d.~C. Bastos, J.~V. N.~A. da~Silva, V.~L. Fabris, V.~d.~S. Uchoa, D.~G. d.~A. Neto, and C.~F. G.~d. Santos.
\newblock Visual question answering: A survey on techniques and common trends in recent literature.
\newblock {\em arXiv preprint arXiv:2305.11033}, 2023.

\bibitem{Dietrich2014Baseball4DAT}
C.~A. Dietrich, D.~Koop, H.~T. Vo, and C.~T. Silva.
\newblock Baseball4d: A tool for baseball game reconstruction \& visualization.
\newblock {\em 2014 IEEE Conference on Visual Analytics Science and Technology (VAST)}, pp. 23--32, 2014.

\bibitem{Fu2023HoopInSightAA}
Y.~Fu and J.~T. Stasko.
\newblock Hoopinsight: Analyzing and comparing basketball shooting performance through visualization.
\newblock {\em IEEE Transactions on Visualization and Computer Graphics}, 30:858--868, 2023.

\bibitem{gershon2001storytelling}
N.~Gershon and W.~Page.
\newblock What storytelling can do for information visualization.
\newblock {\em Communications of the ACM}, 44(8):31--37, 2001.

\bibitem{green2004transportation}
M.~C. Green.
\newblock Transportation into narrative worlds: The role of prior knowledge and perceived realism.
\newblock {\em Discourse processes}, 38(2):247--266, 2004.

\bibitem{javed2012exploring}
W.~Javed and N.~Elmqvist.
\newblock Exploring the design space of composite visualization.
\newblock In {\em 2012 ieee pacific visualization symposium}, pp. 1--8. IEEE, 2012.

\bibitem{kohli2015optimal}
I.~S. Kohli.
\newblock On optimal offensive strategies in basketball.
\newblock {\em arXiv preprint arXiv:1506.06687}, 2015.

\bibitem{kraak2003space}
M.-J. Kraak.
\newblock The space-time cube revisited from a geovisualization perspective.
\newblock In {\em Proc. 21st international cartographic conference}, pp. 1988--1996. Citeseer, 2003.

\bibitem{kriglstein2014pep}
S.~Kriglstein, M.~Pohl, and M.~Smuc.
\newblock Pep up your time machine: Recommendations for the design of information visualizations of time-dependent data.
\newblock {\em Handbook of human centric visualization}, pp. 203--225, 2014.

\bibitem{lewis2020retrieval}
P.~Lewis, E.~Perez, A.~Piktus, F.~Petroni, V.~Karpukhin, N.~Goyal, H.~K{\"u}ttler, M.~Lewis, W.-t. Yih, T.~Rockt{\"a}schel, et~al.
\newblock Retrieval-augmented generation for knowledge-intensive nlp tasks.
\newblock {\em Advances in Neural Information Processing Systems}, 33:9459--9474, 2020.

\bibitem{Lin2023VIRDIM}
T.~Lin, A.~Aouididi, C.~Zhu-Tian, J.~Beyer, H.~Pfister, and J.-H. Wang.
\newblock {VIRD: Immersive Match Video Analysis for High-Performance Badminton Coaching}.
\newblock {\em IEEE Transactions on Visualization and Computer Graphics}, 30:458--468, 2023.

\bibitem{lin2022quest}
T.~Lin, C.~Zhu-Tian, Y.~Yang, D.~Chiappalupi, J.~Beyer, and H.~Pfister.
\newblock The quest for omnioculars: Embedded visualization for augmenting basketball game viewing experiences.
\newblock {\em IEEE transactions on visualization and computer graphics}, 29(1):962--971, 2022.

\bibitem{liu2024visual}
H.~Liu, C.~Li, Q.~Wu, and Y.~J. Lee.
\newblock Visual instruction tuning.
\newblock {\em Advances in neural information processing systems}, 36, 2024.

\bibitem{Losada2016BKVizAB}
A.~G. Losada, R.~Ther{\'o}n, and A.~Benito.
\newblock Bkviz: A basketball visual analysis tool.
\newblock {\em IEEE Computer Graphics and Applications}, 36:58--68, 2016.

\bibitem{mayr2018once}
E.~Mayr and F.~Windhager.
\newblock Once upon a spacetime: Visual storytelling in cognitive and geotemporal information spaces.
\newblock {\em ISPRS International Journal of Geo-Information}, 7(3):96, 2018.

\bibitem{mcintyre2016recognizing}
A.~McIntyre, J.~Brooks, J.~Guttag, and J.~Wiens.
\newblock Recognizing and analyzing ball screen defense in the nba.
\newblock In {\em Proceedings of the MIT sloan sports analytics conference, Boston, MA, USA}, pp. 11--12, 2016.

\bibitem{miall2001shifting}
D.~S. Miall and D.~Kuiken.
\newblock Shifting perspectives: Readers’ feelings and literary response.
\newblock {\em New perspectives on narrative perspective}, pp. 289--301, 2001.

\bibitem{mulcahy2016positioning}
M.~Mulcahy and B.~Gouldthorp.
\newblock Positioning the reader: the effect of narrative point-of-view and familiarity of experience on situation model construction.
\newblock {\em Language and Cognition}, 8(1):96--123, 2016.

\bibitem{o2010development}
H.~L. O'Brien and E.~G. Toms.
\newblock The development and evaluation of a survey to measure user engagement.
\newblock {\em Journal of the American Society for Information Science and Technology}, 61(1):50--69, 2010.

\bibitem{park2023generative}
J.~S. Park, J.~O'Brien, C.~J. Cai, M.~R. Morris, P.~Liang, and M.~S. Bernstein.
\newblock Generative agents: Interactive simulacra of human behavior.
\newblock In {\em Proceedings of the 36th Annual ACM Symposium on User Interface Software and Technology}, pp. 1--22, 2023.

\bibitem{Perin2013SoccerStoriesAK}
C.~Perin, R.~Vuillemot, and J.-D. Fekete.
\newblock Soccerstories: A kick-off for visual soccer analysis.
\newblock {\em IEEE Transactions on Visualization and Computer Graphics}, 19:2506--2515, 2013.

\bibitem{roberts2005exploratory}
J.~C. Roberts.
\newblock Exploratory visualization with multiple linked views.
\newblock In {\em Exploring geovisualization}, pp. 159--180. Elsevier, 2005.

\bibitem{salvador2007toward}
S.~Salvador and P.~Chan.
\newblock Toward accurate dynamic time warping in linear time and space.
\newblock {\em Intelligent Data Analysis}, 11(5):561--580, 2007.

\bibitem{schnotz2005integrated}
W.~Schnotz.
\newblock An integrated model of text and picture comprehension.
\newblock {\em The Cambridge handbook of multimedia learning}, 49(2005):69, 2005.

\bibitem{schroder2023telling}
K.~Schr{\"o}der, W.~Eberhardt, P.~Belavadi, B.~Ajdadilish, N.~van Haften, E.~Overes, T.~Brouns, and A.~C. Valdez.
\newblock Telling stories with data--a systematic review.
\newblock {\em arXiv preprint arXiv:2312.01164}, 2023.

\bibitem{segel2010narrative}
E.~Segel and J.~Heer.
\newblock Narrative visualization: Telling stories with data.
\newblock {\em IEEE transactions on visualization and computer graphics}, 16(6):1139--1148, 2010.

\bibitem{shen2023data}
L.~Shen, Y.~Zhang, H.~Zhang, and Y.~Wang.
\newblock Data player: Automatic generation of data videos with narration-animation interplay.
\newblock {\em IEEE Transactions on Visualization and Computer Graphics}, 2023.

\bibitem{sicilia2019deephoops}
A.~Sicilia, K.~Pelechrinis, and K.~Goldsberry.
\newblock Deephoops: Evaluating micro-actions in basketball using deep feature representations of spatio-temporal data.
\newblock In {\em Proceedings of the 25th ACM SIGKDD International Conference on Knowledge Discovery \& Data Mining}, pp. 2096--2104, 2019.

\bibitem{singh2023optimizing}
A.~Singh.
\newblock Optimizing performance in basketball: A game-theoretic approach to shot percentage distribution in a team.
\newblock {\em arXiv e-prints}, pp. arXiv--2310, 2023.

\bibitem{skinner2017optimal}
B.~Skinner and M.~Goldman.
\newblock Optimal strategy in basketball.
\newblock In {\em Handbook of statistical methods and analyses in sports}, pp. 245--260. Chapman and Hall/CRC, 2017.

\bibitem{Stein2018BringIT}
M.~Stein, H.~Janetzko, A.~Lamprecht, T.~Breitkreutz, P.~Zimmermann, B.~Goldl{\"u}cke, T.~Schreck, G.~L. Andrienko, M.~Grossniklaus, and D.~A. Keim.
\newblock Bring it to the pitch: Combining video and movement data to enhance team sport analysis.
\newblock {\em IEEE Transactions on Visualization and Computer Graphics}, 24:13--22, 2018.

\bibitem{sultanum2023datatales}
N.~Sultanum and A.~Srinivasan.
\newblock Datatales: Investigating the use of large language models for authoring data-driven articles.
\newblock In {\em 2023 IEEE Visualization and Visual Analytics (VIS)}, pp. 231--235. IEEE, 2023.

\bibitem{tian2019use}
C.~Tian, V.~De~Silva, M.~Caine, and S.~Swanson.
\newblock Use of machine learning to automate the identification of basketball strategies using whole team player tracking data.
\newblock {\em Applied Sciences}, 10(1):24, 2019.

\bibitem{tsai2017recognizing}
T.-Y. Tsai, Y.-Y. Lin, H.-Y.~M. Liao, and S.-K. Jeng.
\newblock Recognizing offensive tactics in broadcast basketball videos via key player detection.
\newblock In {\em 2017 IEEE International Conference on Image Processing (ICIP)}, pp. 880--884. IEEE, 2017.

\bibitem{tversky2002animation}
B.~Tversky, J.~B. Morrison, and M.~Betrancourt.
\newblock Animation: can it facilitate?
\newblock {\em International journal of human-computer studies}, 57(4):247--262, 2002.

\bibitem{unkown10glossary}
Unkown.
\newblock Glossary of basketball terms.
\newblock \url{"https://en.wikipedia.org/wiki/Glossary_of_basketball_terms"}, Oct. 2010.
\newblock Accessed on March 14, 2024.

\bibitem{Wang2021TacMinerVT}
J.~Wang, J.~Wu, A.~Cao, Z.~Zhou, H.~Zhang, and Y.~Wu.
\newblock Tac-miner: Visual tactic mining for multiple table tennis matches.
\newblock {\em IEEE Transactions on Visualization and Computer Graphics}, 27:2770--2782, 2021.

\bibitem{Wang2020TacSimurTS}
J.~Wang, K.~Zhao, D.~Deng, A.~Cao, X.~Xie, Z.~Zhou, H.~Zhang, and Y.~Wu.
\newblock Tac-simur: Tactic-based simulative visual analytics of table tennis.
\newblock {\em IEEE Transactions on Visualization and Computer Graphics}, 26:407--417, 2020.

\bibitem{willett2016embedded}
W.~Willett, Y.~Jansen, and P.~Dragicevic.
\newblock Embedded data representations.
\newblock {\em IEEE transactions on visualization and computer graphics}, 23(1):461--470, 2016.

\bibitem{Wu2022OBTrackerVA}
Y.~Wu, D.~Deng, X.~Xie, M.~He, J.~Xu, H.~Zhang, H.~Zhang, and Y.~Wu.
\newblock Obtracker: Visual analytics of off-ball movements in basketball.
\newblock {\em IEEE Transactions on Visualization and Computer Graphics}, 29:929--939, 2022.

\bibitem{Yao2023DesigningFV}
L.~Yao, R.~Vuillemot, A.~Bezerianos, and P.~Isenberg.
\newblock Designing for visualization in motion: Embedding visualizations in swimming videos.
\newblock {\em IEEE Transactions on Visualization and Computer Graphics}, 30:1821--1836, 2023.

\bibitem{yao2022react}
S.~Yao, J.~Zhao, D.~Yu, N.~Du, I.~Shafran, K.~Narasimhan, and Y.~Cao.
\newblock React: Synergizing reasoning and acting in language models.
\newblock {\em arXiv preprint arXiv:2210.03629}, 2022.

\bibitem{zhang2023gpt4roi}
S.~Zhang, P.~Sun, S.~Chen, M.~Xiao, W.~Shao, W.~Zhang, K.~Chen, and P.~Luo.
\newblock Gpt4roi: Instruction tuning large language model on region-of-interest.
\newblock {\em arXiv preprint arXiv:2307.03601}, 2023.

\bibitem{zhao2015data}
Z.~Zhao, R.~Marr, and N.~Elmqvist.
\newblock Data comics: Sequential art for data-driven storytelling.
\newblock {\em tech. report}, 2015.

\bibitem{Zhi2019GameViewsUA}
Q.~Zhi, S.~Lin, P.~T. Sukumar, and R.~A. Metoyer.
\newblock Gameviews: Understanding and supporting data-driven sports storytelling.
\newblock {\em Proceedings of the 2019 CHI Conference on Human Factors in Computing Systems}, 2019.

\bibitem{Zhi2020GameBotAV}
Q.~Zhi and R.~A. Metoyer.
\newblock Gamebot: A visualization-augmented chatbot for sports game.
\newblock {\em Extended Abstracts of the 2020 CHI Conference on Human Factors in Computing Systems}, 2020.

\bibitem{chen2023iball}
C.~Zhu-Tian, Q.~Yang, J.~Shan, T.~Lin, J.~Beyer, H.~Xia, and H.~Pfister.
\newblock {iBall: Augmenting Basketball Videos with Gaze-Moderated Embedded Visualizations}.
\newblock In {\em Proceedings of the 2023 CHI Conference on Human Factors in Computing Systems}, pp. 1--18, 2023.

\bibitem{zhu2022sporthesia}
C.~Zhu-Tian, Q.~Yang, X.~Xie, J.~Beyer, H.~Xia, Y.~Wu, and H.~Pfister.
\newblock Sporthesia: Augmenting sports videos using natural language.
\newblock {\em arXiv e-prints}, pp. arXiv--2209, 2022.

\bibitem{Chen2021Augmenting}
C.~Zhu-Tian, S.~Ye, X.~Chu, H.~Xia, H.~Zhang, H.~Qu, and Y.~Wu.
\newblock {Augmenting Sports Videos with VisCommentator}.
\newblock {\em IEEE Transactions on Visualization and Computer Graphics}, PP:1--1, 2021.

\end{thebibliography}

\clearpage
\section*{A. Action Detection and Tactic Classification Performance}
\subsection*{A.1 Action Detection}
We selected one quarter from each of the GSW vs. CLE and OKC vs. LAC games. Both were 9 minutes long. 
Our two authors watched the videos and manually reviewed, and revised the results extracted by the action detection model. 
The inter-coder agreement level was 75\%, which was calculated using Cohen's kappa. 
Any disagreements were resolved through discussion between the two authors.
Through discussion, they reached a consensus to establish the ground truth. 
% We extracted key actions to explain the tactics. 
The results are that we extracted a total of 117 actions: 59 passes, 15 screens, 14 cuts, and 29 shoots in the GSW vs. CLE game.
Similarly, in the OKC vs. LAC game, we identified 104 actions: 45 passes, 24 screens, 8 cuts, and 27 shoots.
% We manually reviewed all the results. 
% Two authors watched the videos, recording the classification and timing of each action. 
% Through discussion, they reached a consensus to establish the ground truth. 
% Based on this ground truth, the accuracy of action detection is XX.
The overall F1 score was 0.7393, and the confusion matrix is provided in \autoref{tab:action}.
% The differences in F1 score and accuracy arose due to the imbalance in the data. 
We used the manually revised actions from the model in the actual experiments.

\begin{table}[h!]
\centering
\begin{tabular}{|l|l|l|l|l|}
\hline
       & Cut  & Pass & Screen & Shoot \\ \hline
Cut    & 0.5  & 0.36 & 0.0    & 0.14  \\ \hline
Pass   & 0.06 & 0.63 & 0.07   & 0.24  \\ \hline
Screen & 0.08 & 0.22 & 0.48   & 0.22  \\ \hline
Shoot  & 0.0  & 0.17 & 0.08   & 0.75  \\ \hline
\end{tabular}
\caption{A confusion matrix of the action detection}
\label{tab:action}
\end{table}

% Action counts from game1: Counter({'Pass': 78, 'Screen': 63, 'Shoot': 27, 'Cut': 20})
% Action counts from game2: Counter({'Pass': 62, 'Screen': 59, 'Shoot': 13, 'Cut': 10})

% Game 1: {'Cut': 14, 'Pass': 59, 'Screen': 15, 'Shoot': 29} - 117
% Game 2: {'Screen': 24, 'Pass': 45, 'Shoot': 27, 'Cut': 8} - 104
% 0.7504116812871625
% 0.7356073362680265
% 0.7663548220770016
% [[0.5        0.36363636 0.         0.13636364]
%  [0.05645161 0.62903226 0.07258065 0.24193548]
%  [0.08       0.22       0.48       0.22      ]
%  [0.         0.16666667 0.08333333 0.75      ]]
% f1: 0.7393939393939394, acc: 0.5865384615384616
% ['Cut', 'Pass', 'Screen', 'Shoot']

\subsection*{A.2 Tactic Classification}
We built our tactic classification model based on the previous work \cite{tsai2017recognizing}, which includes 134 videos from the NBA 2013-2014 season, covering 10 different offensive tactics. 
The dataset consists of the key players' trajectory coordinates data and their corresponding labels. 
Each trajectory is denoted as $\{(x, y)\}_{i=0}^t$, with $x$ and $y$ are the on-court position, and $t$ marking the clip's end frame. 
Specifically, we leverage the K-Nearest Neighbors (KNN) algorithm to identify the closest match. 
We evaluated our model by following the experimental procedure described in the previous work. 
The study was validated using k-fold cross-validation, where the dataset was divided into 5 subsets.
The overall accuracy was 0.8533, and the confusion matrix is provided in \autoref{tab:slim_table}. The labels represent offensive tactics: F23 (2-3 Flex), EV (Elevator), HK (Hawk), PD (Pin-Down), PT (Princeton), RB (Back-Side Pick and Roll), SP (Side-Pick Slip and Pop), WS (Warrior Single), WV (Weave), and WW (Wing-Wheel). The predicted tactics are given in the columns, with diagonal entries representing correct classifications. The overall accuracy of this classifier is lower than that reported in the previous work \cite{tsai2017recognizing}. However, we believe this level of accuracy is acceptable for providing users with useful explanations of the tactics.

\begin{table}[h!]
    \centering
    \begin{adjustbox}{max width=0.5\textwidth}
    \begin{tabular}{|>{\centering\arraybackslash}p{1cm}|>{\centering\arraybackslash}p{1cm}|>{\centering\arraybackslash}p{1cm}|>{\centering\arraybackslash}p{1cm}|>{\centering\arraybackslash}p{1cm}|>{\centering\arraybackslash}p{1cm}|>{\centering\arraybackslash}p{1cm}|>{\centering\arraybackslash}p{1cm}|>{\centering\arraybackslash}p{1cm}|>{\centering\arraybackslash}p{1cm}|>{\centering\arraybackslash}p{1cm}|}
        \hline
        Accu. & F23  & EV   & HK & PD   & PT   & RB   & SP   & WS   & WV   & WW \\ \hline
        F23   & 0.86 & 0.07 & 0  & 0    & 0    & 0    & 0    & 0.07 & 0    & 0  \\ \hline
        EV    & 0.10 & 0.80 & 0  & 0    & 0.10 & 0    & 0    & 0    & 0    & 0  \\ \hline
        HK    & 0    & 0    & 1  & 0    & 0    & 0    & 0    & 0    & 0    & 0  \\ \hline
        PD    & 0    & 0    & 0  & 0.90 & 0.10 & 0    & 0    & 0    & 0    & 0  \\ \hline
        PT    & 0.17 & 0    & 0  & 0    & 0.60 & 0    & 0.23 & 0    & 0    & 0  \\ \hline
        RB    & 0    & 0    & 0  & 0    & 0    & 0.93 & 0.07 & 0    & 0    & 0  \\ \hline
        SP    & 0    & 0    & 0  & 0    & 0    & 0    & 0.80 & 0.20 & 0    & 0  \\ \hline
        WS    & 0    & 0    & 0  & 0    & 0    & 0    & 0    & 1    & 0    & 0  \\ \hline
        WV    & 0    & 0    & 0  & 0    & 0.18 & 0.07 & 0    & 0.07 & 0.68 & 0  \\ \hline
        WW    & 0    & 0    & 0  & 0    & 0    & 0    & 0    & 0    & 0    & 1  \\ \hline
    \end{tabular}
    \end{adjustbox}
    \caption{A confusion matrix of the tactic classifier. }
    \label{tab:slim_table}
\end{table}

% To gain insight into the average accuracy, the confusion matrix of our approach is given in Table 3, where the predicted tactics are given in columns. Diagonal entries represent correct classification. The confusion matrix indicates that except tactic WS, all the tactics have accuracies higher than 90%.

\section*{B. Prompts examples}
\subsection*{B.1 Prompts for generating overview of tactics}

\subsubsection*{B.1.1 Third-person perspectives}
\begin{lstlisting}
"""
You are the basketball coach who knows basketball tactics. 
The tactic description and actions are provided.        
Please briefly explain the question from casual fans.
When explaining offensive tactics, describe using the attacking players, and when explaining defensive tactics, describe using the defending players.

[PLAYER INFORMATION]
Offense Players: {(*@\textbf{Offensive Players}@*)}
Defense Players: {(*@\textbf{Defensive Players}@*)}

[CONSTRAINT] 
Note that you have to answer integrating tactic description and actions within 2 sentences.

[TACTIC]
{(*@\textbf{Tactic Description}@*)}

[ACTION]
{(*@\textbf{Actions}@*)}
    
Please explain {(*@\textbf{Question}@*)}.
"""
\end{lstlisting}
\subsubsection*{B.1.2 First-person perspectives}
\begin{lstlisting}
"""
You are the basketball coach who knows basketball tactics. 
The response should be in the format of a role-play dialogue between players, and no other text should be added besides the players' conversation.
I would like it to consist of only 2 to 4 conversations between players.
When explaining offensive tactics, describe using the attacking players, and when explaining defensive tactics, describe using the defending players.

[ANSWER FORMAT]
The Answer Format is as follows: 
Stephen Curry: Alright, let's now use the pick and roll tactic. Draymond Green, set a screen for me. Then, I'll make my move. \n\n
Draymond Green: Let's confuse the opponent with our tactic and score. \n\n


[PLAYER INFORMATION]
Offense Players: {(*@\textbf{Offensive Players}@*)}
Defense Players: {(*@\textbf{Defensive Players}@*)}

[CONSTRAINT] 
Note that you have to answer integrating tactic description and actions within 2 sentences.

[TACTIC]
{(*@\textbf{Tactic Description}@*)}

[ACTION]
{(*@\textbf{Actions}@*)}

Please explain {(*@\textbf{Question}@*)}.
"""
\end{lstlisting}

\subsection*{B.2 Prompts for generating action-by-action explanation}
\subsubsection*{B.2.1 Third-person perspectives}
\begin{lstlisting}
"""
The actions represent player's movements such as Cut, Pass, Screen, and Shoot. 

[CONSTRAINT]
1. The format of the answer must be the same as the explanation and must be described in third person within 1 sentence.
2. The names of all players must be accurately written.
3. When explaining offensive tactics, describe using the attacking players, and when explaining defensive tactics, describe using the defending players. Answer only one offensive or defensive tactic for me.
4. All conversations should include the reason why that player is taking such action.
The reasons for cut: create scoring opportunities, disturb the defense, enhance ball movement, space the floor, or implement offensive strategy.
The reasons for pass: Create Better Scoring Opportunities, Control the Pace of the Game, Enhance Team Play, Overcome Tight Defense, or Improve Court Vision and Awareness.
The reasons for screen: Disrupt Defensive Schemes, Force Defensive Adjustments,  Diversify Offensive Strategies, or Creating space for other players.

[PLAYER INFORMATION]
Offense Players: {(*@\textbf{Offensive Players}@*)}
Defense Players: {(*@\textbf{Defensive Players}@*)}

[ANSWER FORMAT]
1. Generate the answer using only conversational responses, without including any action titles or additional explanations.
2. Each conversation will be divided by \n\n and ensure that the Answer Format is as follows:

Action 1. Explanation 1 \n\n Action 2. Explanation 2 \n\n Action 3. Explanation 3 \n\n

[ACTION]
{(*@\textbf{Actions}@*)}

Please explain question "{(*@\textbf{question}@*)}" based on each individual action with reasoning
"""

\end{lstlisting}
\subsubsection*{B.2.2 First-person perspectives}
\begin{lstlisting}
"""
The actions represent player's movements such as Cut, Pass, Screen, and Shoot. 

[CONSTRAINT]
1. The answer format should be conversation between two players if the action is the interaction between two players with first person perspective.
2. In the "Screen", the two players are on different teams, and one player is setting a screen on another. It\'s important not to inform or alert the other player about the screen being set; instead, dialogue should be created that disrupts or expresses confusion.
3. Please ensure that the player name is the full name, and the "Shoot" action results in only one answer from the Shooter. 
4. The response should be in the format of a role-play dialogue between players, and no other text should be added besides the players\' conversation. 
5. When explaining offensive tactics, describe using the attacking players, and when explaining defensive tactics, describe using the defending players. Please explain only one of these in response to the question.
6. All conversations SHOULD include the detailed and complex reason why that player is taking such action within one or two sentences for one conversation by referring to below reasons for each action.
The reasons for cut: create scoring opportunities, disturb the defense, enhance ball movement, space the floor, or implement offensive strategy.
The reasons for pass: Create Better Scoring Opportunities, Control the Pace of the Game, Enhance Team Play, Overcome Tight Defense, or Improve Court Vision and Awareness.
The reasons for screen: Disrupt Defensive Schemes, Force Defensive Adjustments, Diversify Offensive Strategies, or Creating space for other players.

[PLAYER INFORMATION]
Offense Players: {(*@\textbf{Offensive Players}@*)}
Defense Players: {(*@\textbf{Defensive Players}@*)}

[ANSWER FORMAT]
1. Generate the answer using only conversational responses with reasons, without including any action titles or additional explanations.
2. "Pass" and "Screen" should consist of two conversations, while "Shoot" should consist of one conversation.
This example is an answer format of "Pass" and "Screen" actions. 
Example 1) Action: Pass Player 1 -> Player 2 Answer: Player 1: Conversation 1 \n Player 2: Conversation 2 \n\n 
Example 2) Action: Screen Player 3 -> Player 4 Answer: Player 3: Conversation 3 \n Player 4: Conversation 4 \n\n 
This example is an answer format of "Shoot" action. Example) Action: Shoot Player 2 Answer: Player 2: Conversation 5 \n
3. Each conversation will be divided by and ensure that the Answer Format is as follows:
Player 1: Conversation 1 \n Player 2: Conversation 2 \n\n Player 2: Conversation 3 \n Player 3: Conversation 4 \n\n Player 2: Conversation 5 \n
Make sure these \n and \n\n delimiter.

[ACTION]
{(*@\textbf{Actions}@*)}


PLEASE FOLLOW and MAKE SURE [Answer Format]!!!.
Please explain question "{(*@\textbf{question}@*)}" based on each individual action with reasoning
"""

\end{lstlisting}

\subsection*{B.3 Examples of variables}
\begin{lstlisting}
{(*@\textbf{Offensive Players}@*)}:Andre Iguodala, Stephen Curry, Klay Thompson, Draymond Green
{(*@\textbf{Defensive Players}@*)}:Kevin Love, LeBron James\
{(*@\textbf{Tactic Description}@*)}:The down screen consists of a basketball strategy that occurs when one offensive player faces the general direction of the baseline to set a screen on a defender who is guarding a second offensive player. Following that, the second offensive player can then use the down screen to get open for a possible scoring or playmaking opportunity. The pin down screen is a type of down screen that is typically set near or within the lane at an angle towards the basket. The pin down screen is generally utilized to help an offensive player get open, especially to take a mid-range or three-point jump shot near the perimeter areas of the court. The pin down screen could also be used in conjunction with an offensive set, generally known as floppy action or a floppy screen. Essentially, with the floppy action, an offensive player gets in position near or under the basket and then that same offensive player can cut off a single pin down screen set near the side of one particular lane line or a double screen, which is usually a stagger screen, set near the side of the opposite lane line.
{(*@\textbf{Actions}@*)}:
1.  Cut Draymond Green Wing -> Top and Pass Stephen Curry -> Draymond Green\n
2.  Pass Draymond Green -> Stephen Curry\n
3.  Screen Draymond Green -> Kevin Love\n
4.  Cut Andre Iguodala Post -> Wing and Pass Stephen Curry -> Andre Iguodala\n
5.  Screen Draymond Green -> LeBron James\n
6.  Cut Klay Thompson Wing -> Top and Pass Andre Iguodala -> Klay Thompson\n
7.  Shoot Klay Thompson
{(*@\textbf{question}@*)}:
What happened?
What is the offensive tactic here?
Why does Draymond Green pass the ball to Stephen Curry?

\end{lstlisting}

% \appendix % You can use the `hideappendix` class option to skip everything after \appendix

% \section{About Appendices}
% Refer to \cref{sec:appendices_inst} for instructions regarding appendices.

% \section{Troubleshooting}
% \label{appendix:troubleshooting}

% \subsection{ifpdf error}

% If you receive compilation errors along the lines of \texttt{Package ifpdf Error: Name clash, \textbackslash ifpdf is already defined} then please add a new line \verb|\let\ifpdf\relax| right after the \verb|\documentclass[journal]{vgtc}| call.
% Note that your error is due to packages you use that define \verb|\ifpdf| which is obsolete (the result is that \verb|\ifpdf| is defined twice); these packages should be changed to use \verb|ifpdf| package instead.

% \subsection{\texttt{pdfendlink} error}

% Occasionally (for some \LaTeX\ distributions) this hyper-linked bib\TeX\ style may lead to \textbf{compilation errors} (\texttt{pdfendlink ended up in different nesting level ...}) if a reference entry is broken across two pages (due to a bug in \verb|hyperref|).
% In this case, make sure you have the latest version of the \verb|hyperref| package (i.e.\ update your \LaTeX\ installation/packages) or, alternatively, revert back to \verb|\bibliographystyle{abbrv-doi}| (at the expense of removing hyperlinks from the bibliography) and try \verb|\bibliographystyle{abbrv-doi-hyperref}| again after some more editing.

\end{document}